\documentclass[11pt]{article}   \usepackage{ASA_preamble}

\title{Twists and Gorms and Antifields, oh my!}

\author{Alex S.~Arvanitakis\thanks{Email: \texttt{alex.s.arvanitakis@irb.hr}} }
\affil{\small
\textit{Institut Ruđer Bošković, 54 Bijenička street, 10000 Zagreb, Croatia}
}

\date{\today}
\DeclareMathOperator{\rdeg}{Rdeg}
\DeclareMathOperator{\gh}{ghdeg}
\DeclareMathOperator{\tgh}{twghdeg}
\DeclareMathOperator{\pdeg}{pureghdeg}
\DeclareMathOperator{\anti}{antifdeg}

\begin{document}
\begin{titlepage}
\maketitle
\thispagestyle{empty}
\begin{narrower}
\begin{abstract}
    

    I define topological twists of supersymmetric field theories in the case when the supercharges involved obey an ``open'' algebra. Using the Batalin–Vilkovisky field–antifield formalism, I construct twisted theories algorithmically from the supersymmetry data, and explain supersymmetric localisation in terms of anticanonical transformations. I also treat equivariant topological twists and explain how BV observables contain the equivariant cohomology of the space of histories. Some results are generalised to theories with  two topological supercharges --- such as the "balanced" topological field theories of Dijkgraaf and Moore --- using the geometry of ``differential gorms'' of Kochan and \v{S}evera. Finally, I exhibit  examples of these constructions, including a $\mathrm{U}(1)$-equivariant topological B-model.

\end{abstract}
\end{narrower}
\end{titlepage}

\tableofcontents

\section{Introduction and summary}
In the interests of getting results out --- as opposed to growing large \texttt{.pdf}s on the cloud --- I shall assume anyone reading this has a passing familiarity with supersymmetric (SUSY) localisation, topological twists of supersymmetric theories, and the Batalin-Vilkovisky (BV) formalism, so that I need not summarise or motivate these topics.

A big motivation for this work is to treat the supersymmetric localisation of \emph{on-shell SUSY} theories. (This problem has also been considered in the recent works  \cite{Losev:2023gsq,Lysov:2024lge,cattaneo2025equivariant} and also in the older references \cite{nersesyan1993equivariant,Nersessian:1993me}.) The simplest such theory has a single real supercharge $Q$ that is nilpotent modulo equations of motion:
$$Q^2F= G\frac{\delta S}{\delta \phi}\qquad (F, G \text{ being some functionals; $S$ being the action.})\,.$$ ($Q$ may also be said to be nilpotent up to homotopy of the Koszul differential $\delta$.)
Hitherto, which is to say outside of the BV formalism, these have been treated through the ad hoc introduction of auxiliary fields such that $Q$ squares to zero off-shell. (In other words, $Q^2=0$ at the level of cochains rather than cohomology.) I give an alternative, less ad-hoc approach, whose crucial component is the \emph{BV realisation of the topological twist} by the supercharge $Q$. Such a twist is the choice of a $\mathbb Z$-grading; it comes from R-charge in supersymmetric gauge theory. Given the twist, standard methods of homological perturbation theory familiar to BV practicioners produce a \emph{new} lagrangian and a new solution to the master equation, whose antifield-BRST cohomology encodes $Q$-closed (i.e.~BPS) observables modulo equations of motion.

(At this point I should clarify that I take the phrase \emph{topological twist} to mean the specific construction where the physical operators of the twisted theory correspond to the BPS operators of the original theory. Although this does not clash with common usage in the literature, it is important to point out that the constructions in this paper do not necessarily imply that the twisted theories are independent of metric or other moduli. That being understood, and in keeping with the \emph{topological twist} nomenclature, I will sometimes call the supercharges involved \emph{topological supercharges}.)

Armed with a solution of the master equation corresponding to the topologically twisted theory, localisation is realised by virtue of the independence of the BV path integral with respect to deformations of the lagrangian submanifold that the integral is defined against. The mechanism is identical the one that demonstrates the gauge-fixing independence of correlators in the BV formalism. Unfortunately, as was also observed in the work of Lysov and Losev recently \cite{Lysov:2024lge}, the localisation argument cannot be made independently of the details of the theory and in particular of the right-hand side of $Q^2=\dots$. Regardless, in actual examples there is an obvious choice of a family of lagrangian submanifolds that realises localisation. This situation is not too dissimilar from the situation with gauge-fixing, whose details also depend on the model in question.

The BV realisation of topological twists is perhaps the key new idea in this work. I employ it to study the more complicated scenario where $Q$ or a collection of supercharges do not anticommute on-shell but instead give rise to one or more infinitesimal symmetries. I shall call the corresponding twisted theories \emph{equivariantly twisted}, and will distinguish between two types:
\begin{itemize}
    \item \textbf{DM-type equivariant twist:} Here we have a Lie group with a Lie algebra $\frak g$ acting infinitesimally through generators $B_a$ which are $Q$-exact on-shell. ($Q$ remains nilpotent on-shell.) Equivalently, the Cartan calculus is realised on the BV manifold only up to homotopy. The DM-type equivariantly twisted theory then realises the Kalkman (`BRST') model of equivariant cohomology \cite{Kalkman:1993zp} of the equation of motion locus.  DM-type twisted theories require a new definition of an equivariant BV formalism and new definitions for \emph{equivariant} classical and quantum master equations relative to the recent approach of references \cite{Bonechi:2019dqk,cattaneo2025equivariant} and I spend some time making sure these definitions are sensible. (See also \cite{Mikhailov:2020lzo,Mikhailov:2022fol,Nersessian:1995yt,Getzler:2018sbh} for discussions of equivariant BV and \cite{Elliott:2018cbx,Borsten:2025hrn} for complementary perspectives on topological twists.) 
    
    \item \textbf{BT-type equivariant twist:} In this situation, the supercharge $Q$ is not nilpotent but squares to a symmetry on-shell, specifically in infinitesimal gauge symmetry with field-dependent gauge parameter. (Although $Q$ is then called ``equivariantly nilpotent'' in some of the literature, this scenario is not obviously connected to equivariant cohomology.) BT-type equivariantly twisted theories obey an unmodified master equation.
\end{itemize}
We shall see later that the types are not quite mutually exclusive.

I also discuss the generalisation appropriate for SUSY theories with multiple topological supercharges, focussing on the case with exactly two such. (Considerations of SUSY algebra show that one should expect an $N=2k$ theory to admit $k$ topological supercharges, so that more than two topological supercharges require a very high amount of supersymmetry.) Topological field theories with two supercharges were introduced by Blau and Thompson \cite{Blau:1991bn,Blau:1991jg} and further studied by Dijkgraaf and Moore \cite{Dijkgraaf:1996tz} where they were christened ``balanced'' topological field theories. A point the latter authors made was that the two supercharges $Q_1$ and $Q_2$ should be thought of as a pair of de Rham differentials. I develop that line of thinking using the geometry of ``differential gorms'' which was articulated by Kochan and \v{S}evera \cite{kochan2003differential,severa2006differential}. The gorm perspective is especially relevant for the DM-type equivariant generalisation, but I also present the BT-type one.

Finally, I discuss examples of theories that fit into these frameworks, treating SUSY quantum mechanics and topological sigma models in some detail, while providing some Yang-Mills-type examples of two-supercharge equivariant twists. The $\mathrm{U}(1)$-equivariant topological B-model in two dimensions that is constructed in section \ref{sec:equivariantBmodel} appears to be novel.

\section{A homological perturbation lemma}

I will use the following result very many times in the sequel, so it deserves its own subsection. 
\begin{theorem}[8.3 \cite{henneaux1992quantization}]
Let $\delta$ and $\mathcal D$ be two odd left derivations acting on some graded-commutative algebra $A$. The algebra is bigraded with $\mathbb Z$-gradings $\gh$ and $\rdeg$ such that $\gh+\rdeg \mod 2$ equals Grassmann parity. The derivations have the degree assignments $\gh \delta= \rdeg\mathcal D=1$ and $\gh \mathcal D=\rdeg \delta=0$ and obey the identities
\be
\delta^2=0\,,\quad [\delta,\mathcal D]=0\,,\quad \mathcal D^2=-[\delta,s_1]
\ee
for $s_1$ some odd derivation of appropriate degrees. Finally, $\delta$ is required to be \emph{acyclic}: $H^{n\neq 0}(\delta)=0$.

Then:
\begin{itemize}
   \item[(a)] There exist $s_2,s_3,\dots$ of $\gh s_n=n$ such that the odd left derivation
   \be
   \label{pertlemmaderivationS}
   s \equiv \delta+d+s_1+\cdots
   \ee
   has  degree 1 in $ \gh+\rdeg$ and is nilpotent: $s^2=0$.
   \item[(b)] Any such $s$ has cohomology the ``cohomology of $\mathcal D$-modulo-$\delta$'':
   \be
   H^\bullet(s)\cong H^\bullet\big(\mathcal D| H^0(\delta)\big)\,,
   \ee
   where classes in $H^n\big(\mathcal D| H^0(\delta)\big)$ are defined through the following equivalence relation $\sim$:
   \be
   \label{cohomologywithvalues:Def}
   \begin{split}
      x\in  A\,,\qquad \gh x=0\,,\qquad \rdeg x=n\,,\quad \delta x=0\,, \\ \mathcal D x=\delta y\,,\qquad
      x\sim x + \mathcal D z + \delta z'\,.
   \end{split}
   \ee
\end{itemize}
\end{theorem}
In general $\delta$ will be the Koszul differential associated to the equation of motion locus, or the Koszul-Tate differential in certain subsections where gauge symmetry must be considered, while $\mathcal D$ will contain the topological supercharges. The algebra $A$ will contain the ring of functions $\cin\equiv C^\infty(\mathcal M)$ of a  ($-1$)-symplectic manifold $\mathcal M$. and $\mathcal M$ will be the shifted cotangent bundle $T^\star[-1] M$ of a (compact) supermanifold $M$. In all examples, $M$ will admit the $\rdeg$ grading, which is then lifted to $T^\star[-1] M$ in the obvious way, while $\gh$ will denote the fibre degree of this cotangent bundle.

\section{A single supercharge}
\subsection{Topological twists}
\label{sec:topologicaltwist}
To begin with, consider a classical theory without gauge symmetries. In this familiar situation, the proper solution to the classical BV master equation has antifield number zero, so I denote it as $S_0$. (This functional is the ``original action'' which is completed to a proper solution $S_\mathrm{BV}=S_0+S_1+\cdots$ of the master equation by adding terms involving antifields whenever $S_0$ admits gauge symmetries.) Moreover the Koszul differential $\delta\equiv (S_0,\bullet)$ --- where $(\bullet,\bullet)$ is the antibracket --- is acyclic outside antifield number zero ($H_{k\neq 0}(\delta)=0$). Functionals which are $\delta$-exact vanish whenever the equations of motions hold (i.e.~they "vanish on-shell") and so the homology of $\delta$ is identified with the space of functionals of \emph{on-shell field configurations}. These are precisely the classical observables. 

(For purposes of aesthetics it is convenient to speak of \emph{ghost number} $\gh$ instead of antifield number, so that $\gh \delta = +1$ defines a cochain complex rather than a chain complex and a cohomology rather than a homology. This accords with the typical convention in the BV formalism, where the (total) ghost number is minus the antifield number whenever there are no pureghosts. I am in that case except wherever I consider gauge symmetries.)

Let us assume also that exists a Grassmann-\emph{odd} ghost number zero vector field $Q$ with Grassmann-even ghost number $-1$ hamiltonian $\Theta$:
\be
Q=(\Theta,\bullet)\,.
\ee 
This means that ghost number and Grassmann parity are uncorrelated; I still use the sign conventions of \cite{Arvanitakis:2018cyo,Arvanitakis:2021wkt} albeit by replacing the degree there with the total parity ($\gh$ mod 2 + fermion). If $Q$ is a (super)symmetry of the action $S_0$ which is nilpotent on-shell, it satisfies
\be
\label{Qconditions}
[Q,\delta]=0\,,\qquad \mathrm{and}\qquad Q^2=-[\delta,\alpha]
\ee
by definition for some odd derivation $\alpha$ with $\gh \alpha=-1$.

The conditions \eqref{Qconditions} are instantly recognisable to BV practicioners or other enjoyers of homological algebra as two of the assumptions underlying the perturbation lemma for the perturbation of $\delta$ by $Q$, specifically the version employed by Henneaux and Teitelboim in their textbook \cite[Theorem 8.3 (a)]{henneaux1992quantization} on the BV formalism. (The lemma was quoted in the previous section.) The remaining assumption is the existence of a grading, which I will call $\rdeg$, under which
\be
\rdeg Q=+1,\qquad \rdeg \delta=0\,.
\ee
With these assumptions from \eqref{pertlemmaderivationS} we find a derivation
\be
s\equiv \delta + Q +\alpha +\cdots
\ee
which is nilpotent ($s^2=0$) and is of degree $\tgh s=+1$ in  \emph{twisted ghost number}
\be
\tgh\equiv \gh + \rdeg\,.
\ee

To translate back into physics, I further assume that
\be
\rdeg\omega_\mathrm{BV}=0\,.
\ee
(This may be arranged without loss of generality whenever the BV manifold is a shifted cotangent bundle.) Then $s$ is a hamiltonian vector field. Since $s^2=0$ its hamiltonian $S_\mathrm{tBV}$ satisfies the classical master equation:
\be
(S_\mathrm{tBV},S_\mathrm{tBV})=0\,.
\ee

The physical interpretation of $S_\mathrm{tBV}$ is that of the BV master action of the \emph{topologically twisted theory} (by the supersymmetry $Q$). Indeed, the cohomology of its antifield-BRST differential $s$ is calculated via eq.~\eqref{cohomologywithvalues:Def} of the perturbation lemma which entails that the observables that the BV formalism calculates for $S_\mathrm{tBV}$ are precisely the space of $Q$-invariant functionals on-shell. This is the space of observables of the topologically twisted theory.

\subsection{Equivariant topological twists of type DM}
\label{sec:equivariant}
Let us now consider the case where $Q$ squares to zero on-shell but there are also symmetries of the theory which are $Q$-exact. These symmetries are encoded in a Lie algebra $\frak g$ which must therefore act on the BV manifold. The equivariant twist construction will work whenever some of these symmetries can be assembled as to resemble the operations of the Cartan calculus of the \emph{Weil algebra} $W(\frak g)$ associated to $\frak g$, up to homotopy.

\paragraph{The Weil algebra and the Cartan calculus.} This is a lightning summary to establish notation and conventions; I refer to \cite{Kalkman:1993zp} for details. I define the Weil algebra $W(\frak g)$ associated to the Lie algebra $\frak g$ as the differential graded commutative algebra (dgca) generated, over $\mathbb R$, by
\begin{itemize}
    \item $\kappa^a$ (of degree $\rdeg=1$, anticommuting), and 
    \item $u^a$ (of degree $\rdeg =2$, commuting),
\end{itemize}
where $a$ is a Lie algebra index in some basis $\{T_a\}$ for $\frak g$, with structure constants defined via
\be
[T_a,T_b]=f_{ab}{}^c T_c\,.
\ee
The Weil algebra is is equipped with the following (left) differential, called the Weil differential:
\begin{equation}
    d_W\kappa^a= \frac{1}{2} f_{bc}{}^a \kappa^b\kappa^c + u^a\,,\qquad d_Wu^a=f_{bc}{}^a \kappa^b u^c
\end{equation}
of $\rdeg d_W=+1$, which squares to zero, $d_W^2=0$, and is \textbf{acyclic}: $H^\bullet(d_W)\cong \mathbb R$. (One could also work over $\mathbb C$.) On this space one  may realise the \textbf{Cartan calculus} ``interior product'' and ``Lie derivative'' operators, namely $\iota_a\equiv \pd/\pd \kappa^a$ and $L_a\equiv [d_W,\iota_a]$. Note that here and henceforth \textbf{all commutators are graded}, so that in particular the last expression for $L_a$  is an anticommutator.

\paragraph{A BV construction for the ``equivariant twist''.}
I again start from a classical theory with action $S_0$ and no gauge symmetries, so that $\delta\equiv (S_0,\bullet)$ is acyclic, as before. $S_0$ will be an element of the ring of functions (or functionals, for applications to field theory) $C^\infty(\mathcal M)$ where $\mathcal M$ is the graded supermanifold that is equipped with the odd symplectic structure $\omega_\mathrm{BV}$.

Call the graded ring $C^\infty(\mathcal M)$ of functions on $\mathcal M$ just $C^\infty$ for short. Then I postulate the following objects:
\begin{enumerate}
    \item the existence of a $\mathbb Z$ grading $\rdeg$ on $C^\infty$ such that $\rdeg \omega_{\mathrm{BV}}=0$, as before;
    \item a (left) odd derivation $ Q$ on $C^\infty$ with $\gh=0,\rdeg=+1$, as before;
    \item further odd derivations $ \hat Q_a$ on $C^\infty$ with $\gh=0,\rdeg=-1$;
    \item and some even derivations $ B_a$ on $C^\infty$ with $\gh=0,\rdeg=0$.
\end{enumerate}
I assume all three kinds of derivations are \textbf{hamiltonian}: if they preserve the symplectic form then this is automatic for degree ($\gh$) reasons. Note that whenever $\mathcal M$ is an odd tangent bundle $T^\star[-1]M$ and $Q,\hat Q_a,B_a$ are defined as vector fields on $M$ (which is a supermanifold) their lifts to $T^\star[-1]M$ are automatically hamiltonian; I will not distinguish  between $Q,\hat Q_a,B_a$ and their lifts, notationally-speaking.

The $B_a$ will be the vector fields realising the action of the Lie algebra $\frak g$ and will correspond to the $L_a$ ``Lie derivative'' operators of the Cartan calculus of the Weil algebra; the $\hat Q_a$ will correspond to the $\iota_a$ ``interior product'' operators; and $Q$ will correspond to $d_W$. To deduce the identities they must satisfy, let us write down what should be called the \emph{Kalkman equivariant differential} up to homotopy:
\be
\label{KalkmanD}
\mathcal D\equiv   Q + d_W - \kappa^a B_a +u^a \hat Q_a\ee
Note that this is defined to act on $C^\infty\otimes W(\frak g)$. The relative factors are chosen by convention. (Whenever $\mathcal D^2=0$ this is recognised as the differential defining the Kalkman or ``BRST'' model of equivariant cohomology \cite{Kalkman:1993zp}.) I then calculate
\be
\begin{split}
    \mathcal D^2= Q^2+\tfrac{1}{2}\kappa^a\kappa^b \big([B_a,B_b]-f_{ab}{}^c B_c\big) + \tfrac{1}{2} u^a u^b [\hat Q_a,\hat Q_b] \\+\kappa^a u^b \big( f_{ab}{}^c \hat Q^c - [B_a, \hat Q_b]\big) + u^a\big([Q,\hat Q_a]-B_a\big) + \kappa^a [Q,B_a]\,.
\end{split}
\ee

I can now define the equivariantly twisted theory by analogy to the situation of the preceding subsection. The key postulate is that there exists a derivation $s_1$ of appropriate degrees with
\be
\mathcal D^2=-[\delta,s_1]\,.
\ee
Upon expanding $s_1$ in terms of $W(\frak g)$ generators, which gives 
\be
s_1=\alpha - \beta_a \kappa^a + \gamma_a u^a + \frac{1}{2}\epsilon_{ab}\kappa^a\kappa^b - \zeta_{ab}\kappa^a u^b +\frac{1}{2}\eta_{ab}u^au^b
\ee
where $\alpha,\beta_a,\gamma_a,\cdots,\eta_{ab}$ are all derivations on $C^\infty$, I obtain the identities
\begin{subequations}
    \label{WeilAlgebraOnshell_alternativesigns}
    \begin{align}
        Q^2&=-[\delta,\alpha]\,,\\ \label{identityB}
        [Q,B_a]&= -[\delta,\beta_a]\,,\\ \label{identityC}
        [B_a,B_b] - f_{ab}{}^c B_c &=-[\delta,\epsilon_{ab}]\,,\\
        [Q,\hat Q_a]-B_a&=-[\delta,\gamma_a]\,,\\ \label{identityE}
          f_{bc}{}^a \hat Q_a- [B_b,\hat Q_c] &= -[\delta,\zeta_{bc}]\,,\\
        [\hat Q_a,\hat Q_b] &= -[\delta,\eta_{ab}]\,.
    \end{align}    
\end{subequations}
Of these, identities \eqref{identityB} and \eqref{identityC} are dependent on the others, and we have the relations $\beta_a=[Q,\gamma_a] +[\hat Q_a,\alpha]$ and
\be
\epsilon_{bc}= f_{bc}{}^a \gamma_a + [Q,\zeta_{bc}] -[\beta_b,\hat Q_c]-[B_b,\gamma_c]
\ee up to $[\delta,\bullet]$-closed derivations.

The identities in \eqref{WeilAlgebraOnshell_alternativesigns} are equivalent to $\mathcal D^2=-[\delta,s_1]$. They are also recognised as the identities of the Cartan calculus, albeit realised only modulo equations of motion. Assuming also $[\delta,\mathcal D]=0$ (i.e.~$Q,\hat Q_a,B_a$ are symmetries of the equations of motion) the perturbation lemma implies there exists an odd (left) differential
\be
\label{eq:sDifferential}
s\equiv \delta + \mathcal D+s_1+\dots
\ee
of degree 1 in the ``twisted'' grading $\tgh \equiv\rdeg + \gh$. A critical technical point here is the following: since the proof of the perturbation lemma relies on the acyclicity of $\delta$ --- a derivation on $C^\infty$ --- and $\mathcal D$ is a $W(\frak g)$-linear combination of derivations on $C^\infty$, the omitted terms $(\dots)$ above are necessarily $W(\frak g)$-linear, which is to say that there are no derivatives of the form $\pd/\pd c^a$ or $\pd/\pd u^a$ in $s$ arising from the use of the perturbation lemma. It follows that $s-d_W$ is a hamiltonian vector field.

Therefore there exists an $S_\mathrm{tBV}\in C^\infty\otimes W(\frak g)$ of twisted ghost number $\tgh S_\mathrm{tBV}=0$ which satisfies $s=(S_\mathrm{tBV},\bullet)+ d_W$. Since $s^2=0$, $S_\mathrm{tBV}$ obeys the following \textbf{equivariant classical master equation}
\be
\label{eCME}
(S_\mathrm{tBV},\,S_\mathrm{tBV}) + 2 d_W S_\mathrm{tBV}=0\,.
\ee

\paragraph{The equivariant quantum master equation. Integration.}
This obvious generalisation is obtained from the calculation (where I renamed $S_\mathrm{tBV}$ to $S_{\mathrm{BV}}$)
\be
0=(\hbar\Delta + d_W)\exp(S_\mathrm{BV}/\hbar)=\Big(\hbar \Delta S_\mathrm{BV} +\frac{1}{2} (S_\mathrm{BV},S_\mathrm{BV}) + d_W S_\mathrm{BV}\Big)\hbar^{-1} e^{S_\mathrm{BV}/\hbar}
\ee
so that $(\hbar\Delta + d_W)\exp(S_\mathrm{BV}/\hbar)=0$ if and only if
\be
\label{eQME}
\hbar \Delta S_\mathrm{BV} +\frac{1}{2} (S_\mathrm{BV},S_\mathrm{BV}) + d_W S_\mathrm{BV}\,.
\ee
 As with the usual quantum master equation, $S_\mathrm{BV}$ is assumed to be a power series in the formal parameter $\hbar$ with coefficients in $C^\infty\otimes W(\frak g)$. Its $\hbar$-independent term therefore satisfies the equivariant classical master equation from before. $\Delta$ is the BV laplacian appropriate to the odd symplectic manifold $\mathcal M$ and does not ``talk'' to $W(\frak g)$. Therefore, the operator $\hbar \Delta +d_W$ squares to zero.

Equation \eqref{eQME} has appeared  in the literature before --- e.g.~in the discussion of homotopy-algebraic and combinatorial aspects of the path integral in the work \cite{Braun:2017ikg} --- and defines the quantum master equation for \emph{differential} BV algebras\footnote{There are multiple definitions. One relevant definition is mentioned in a remark in the work \cite{Barannikov:1997dwc} which is the earliest reference I can find.}. However \eqref{eQME} seems to have not been considered in the context of equivariant extensions of the BV formalism. In particular it is inequivalent to the recent proposal in the work \cite{Bonechi:2019dqk}.

Now let us define integrals of quantities $f$ satisfying
\be
(\hbar \Delta + d_W) f=0\,.
\ee
Here $f$ is allowed to be an element in $C^\infty\otimes W(\frak g)$ with coefficients in the ring of formal power series $\mathbb R[\hbar]$. To avoid superfluous analytic issues in the following, assume e.g.~that the body $M$ of $\mathcal M$ is compact. 
\begin{definition}
    \label{defIntegral}
    (I implicitly fix a reference half-density in order to define $\Delta$ on functions.) For any lagrangian submanifold $\mathcal L$ of $\mathcal M$ I define the integral map 
    \be
    I: (C^\infty \otimes W(\frak g))_\text{closed} \to H^\bullet(d_W)
    \ee as the cohomology class of
    \be
    \int_\mathcal L f
    \ee
    where $f\in (C^\infty \otimes W(\frak g))_\text{closed}\iff f\in C^\infty \otimes W(\frak g)$ and $(\hbar \Delta + d_W)f=0\,$. The expression $\int_\mathcal L f$ is defined by linearity on homogeneous elements as
    \be
    \int_{\mathcal L}( \alpha\otimes f_\mathcal{M})\equiv \alpha \int_{\mathcal L} f_\mathcal{M}
    \ee
    where $\alpha \in W(\frak g)$, and where $f_\mathcal M$ is an element of $C^\infty$ so that $\int_{\mathcal L} f_\mathcal{M}$ is a usual BV integral.
\end{definition}
Notice that the integral is automatically $d_W$ closed so that the definition makes sense:
\be
d_W \int_\mathcal L f=\int_\mathcal L d_W f=-\hbar \int_{\mathcal L}  \Delta f=0\,.
\ee
Moreover we have \textbf{invariance under perturbations of the lagrangian} $\mathcal L$ i.e.~independence from the gauge fixing. Using the Weinstein lagrangian neighbourhood theorem, which says $\mathcal M\cong T^\star[-1]\mathcal L$ close to $\mathcal L$, we calculate as usual that nearby lagrangians are  parameterised by functionals $\Psi\in C^\infty$ of $\tgh=-1$ and
\be
\delta\int_\mathcal L f\equiv \int_{\delta\mathcal L} f\propto \int_\mathcal L (\Delta f)  \Psi=-h^{-1} d_W \Big( \int_\mathcal L  f \Psi\Big) \,;
\ee
the last expression is manifestly exact in $d_W$ cohomology, which proves the claim. 

\paragraph{Remark.} Since the cohomology of $d_W$ is isomorphic to numbers $\mathbb R$ the image of this integral is not the $G$-invariant polynomials $S(\frak g^\star)^G$. This is in contrast with the equivariant pushforward of equivariant differential forms. (Note, however, that BV observables do include equivariant differential forms.) The rationale for this definition is twofold: first, to establish that BV gauge fixing works as usual in the twisted theory; second, to arrange that correlation functions in the twisted theory are number-valued. More precisely, any correlator that vanishes upon setting $\kappa^a=u^a=0$ is $d_W$-exact and thus unphysical.

If $\mathcal O$ is a quantum BV observable, it must satisfy $(\hbar \Delta +d_W)\big(\mathcal O\exp(S_\text{BV}/\hbar)\big)=0$ in order to define a good integrand in the sense of Definition \ref{defIntegral}. Observables in the twisted theory are thus defined to satisfy
\be
\label{equivariantquantumBVObservable}
\hbar \Delta\mathcal O +(S_\text{BV},\mathcal O) + d_W\mathcal O=0\,.
\ee
I will have more to say about observables in the next subsection.

\paragraph{Homotopies between solutions.}
Let $S_{0}$ and $S_1\in C^\infty\otimes W(\frak g)$ (with coefficients in formal power series $\mathbb R[\hbar]$) solve the equivariant quantum master equation \eqref{eQME}. A homotopy between two such solutions  is defined in the standard way: namely by introducing $T(t,\dr t)\in  C^\infty\otimes W(\frak g)\otimes \Omega^\bullet([0,1])$, where $\Omega^\bullet([0,1])$ are differential forms on the closed interval $[0,1]$ of real numbers, such that  $T$ solves
\be
\label{eq:homotopyQME}
(\hbar \Delta + d_W + \mathrm{d})\exp(T/\hbar)=0 \iff \hbar \Delta T +\frac{1}{2} (T,T) + d_W T + \mathrm{d} T=0
\ee
with $\mathrm{d}$ the de Rham differential on $[0,1]$, and the boundary conditions $T(0,0)=S_0$, $T(1,0)=S_1$. Upon expanding $T(t,\dr t)=S_t   +\dr t \Psi_t$, where $\Psi_t$ is odd of degree $-1$, the above master equation is equivalent to the following two equations, valid for all values of $t\in [0,1]$:
\begin{subequations}
\begin{align}
    \hbar \Delta S_t +\frac{1}{2} (S_t,S_t) + d_W S_t =0 \,,\\
    \frac{\dr S_t}{\dr t}= (S_t,\Psi_t) +d_W \Psi_t  + \hbar \Delta \Psi_t\,,
\end{align}
\end{subequations}
When $\Psi_t$ is constant in $t$ and independent of $\kappa,u$ in $W(\frak g)$, this reduces to the standard formula for the canonical transformation of the BV master action.

Homotopies between observables $\mathcal O_{0,1}$ are defined similarly, in terms of $\mathcal O(t,\dr t)\equiv \mathcal O_t +\dr t \mathcal N_t$. Demanding that $(\hbar \Delta +d_W +\dr)\big(\exp(T/\hbar)\mathcal O\big)=0$ yields 
\be
\hbar \Delta\mathcal O_t +(S_t,\mathcal O_t) + d_W\mathcal O_t=0
\ee
and
\be
\frac{\dr \mathcal O_t}{\dr t}=-(\Psi_t,\mathcal O_t) + \hbar \Delta \mathcal N_t + d_W \mathcal N_t + (S_t,\mathcal N_t)\,.
\ee
The last equation describes the simultaneous effect of a homotopy in the BV master action as well as a homotopy in the observable itself, the latter being generated by $\mathcal N_t$. What is geometrically counter-intuitive is that in this formulation of equivalences of the equivariant quantum master equation, we may perform ``canonical transformations'' generated by $\Psi$'s which depend on $W(\frak g)$ variables.

Homotopies leave expectation values invariant: for any lagrangian $\mathcal L$,
\be
\int_{\mathcal L}  e^{S_1/\hbar}\mathcal O_1 -\int_{\mathcal L}  e^{S_0/\hbar}\mathcal O_0=\int_0^1 \dr t\; \frac{\dr}{\dr t}\int_{\mathcal L} \Big( e^{T/\hbar}\mathcal O\Big)=0
\ee
where I used $(\hbar \Delta +d_W +\dr)\big(\exp(T/\hbar)\mathcal O\big)=0$ for the last equality.

\subsubsection{On equivariant observables and a ``Cartan-style'' cohomology}
\label{ObservablesSection}
Consider semiclassical BV observables: these are cohomology classes of observables $\mathcal O\in C^\infty\otimes W(\frak g)$ that solve \eqref{equivariantquantumBVObservable} to zero order in $\hbar$. Equivalently,
\be
(S, \mathcal O ) + d_W \mathcal O =0
\ee
with $S$ a solution to the equivariant classical master equation \eqref{eCME}. (In other words semiclassical observables are cohomology classes for the  differential $s\equiv (S,\bullet) + d_W$.) In the situation where the solution to \eqref{eCME} was constructed via homological perturbation, the observable $\mathcal O$ is constructed as an expansion in antifields
\be
\mathcal O=\mathcal O_0 +\mathcal O_1+\cdots
\ee
where $\mathcal O_0$ does not involve antifields and is constrained to obey
\be
\mathcal D \mathcal O_0 =-\delta \mathcal O_1
\ee
for $\mathcal D$ the Kalkman operator from \eqref{KalkmanD} and some $\mathcal O_1 \in C^\infty\otimes W(\frak g)$ of appropriate degrees. Conversely, given such $\mathcal O_0$ and $\mathcal O_1$ we may construct a BV observable $\mathcal O$ obeying $s\mathcal O=0$ as an expansion in antifields, and nontrivial BV observables (in the cohomology $H^\bullet(s)$) correspond to nontrivial classes in the equivalence relation $\mathcal O_0\sim \mathcal O_0 + \mathcal D z +\delta w$.

Counterintuitively, perhaps, such $\mathcal O_0$ need not, in general, arise from observables in the untwisted theory which are invariant (\emph{modulo} $ \delta$) under the symmetry generators $B_a$ and the operators $\hat Q_a$! The reason is that the equivariant Kalkman differential up to homotopy $\mathcal D$ \eqref{KalkmanD} is equivalent to the sum $Q+d_W$, as in the ``strict'' case \cite{Kalkman:1993zp}. Indeed, using the Cartan calculus identities \eqref{WeilAlgebraOnshell_alternativesigns} one calculates the Mathai-Quillen like isomorphism
\be
\exp(\kappa^a \hat Q_a) \mathcal D \exp(-\kappa^a \hat Q_a)= Q + d_W
\ee
where I have omitted all terms of the form $[\delta,\bullet]$. Since $H(d_W)\cong\mathbb R$, we see that the cohomology of $\mathcal D$ on $\cin\otimes W(\frak g)$ is isomorphic to the cohomology of $Q$ on $\cin$. Thus, the semiclassical BV observables of the equivariant theory are the same as those of the non-equivariant theory (\emph{modulo} $\delta$).


So, as in the ``strict'' case, we have found that the cohomology of the Kalkman differential does not capture equivariant cohomology. We can remedy this by defining a special class of \textbf{equivariant observables}: $\mathcal O=\mathcal O_0 + \mathcal O_1 +\cdots$ represents a classical equivariant observable if and only if $s\mathcal O=0$ and there exist degree-appropriate $y_a$, $z_a$ such that
\be
\label{eq:CartansubringConditions}
\frac{\pd}{\pd \kappa^a}\mathcal O_0=-\delta y_a\,,\quad \big(B_a - f_{ab}{}^c u^b\frac{\pd}{\pd u^c}\big)\mathcal O_0=-\delta z_a\,.
\ee
In other words, $\mathcal O_0$ defines a ``basic form'' for the Kalkman equivariant cohomology on-shell. A key  point is that these conditions are \emph{not preserved} under the equivalence relation $\mathcal O\sim \mathcal O + s \mathcal P$ \emph{unless} we impose conditions \eqref{eq:CartansubringConditions} on $\mathcal P_0$ since
$\mathcal O_0 \sim \mathcal O_0 + \mathcal D \mathcal P_0 + \delta\mathcal P_1
$. Therefore the correct equivalence relation for equivariant observables is finer compared to the one one would naturally guess in the BV formalism.

To express the equivalence relation for equivariant observables in a less awkward way we may consider the subring 
\be
\begin{split}
    R_\text{Cartan}\equiv \big\{\mathcal O \in \cin\otimes W(\frak g) \; | \\ \exists y^{\mathcal O}_a, z^{\mathcal O}_a \in \cin\otimes W(\frak g): \frac{\pd}{\pd \kappa^a}\mathcal O_0=-\delta y^{\mathcal O}_a\,,\; \quad (B_a- f_{ab}{}^c u^b\tfrac{\pd}{\pd u^c} )\mathcal O_0=-\delta z^{\mathcal O}_a \big\}
\end{split}\ee
of $\cin\otimes W(\frak g)$ where again $\mathcal O_0$ is the antifield-free term in $\mathcal O$; this is a subring because $\delta \mathcal O_0=0$ for any $\mathcal O$. This ring may be thought of as housing the Cartan complex for equivariant cohomology (up to homotopy of $\delta$). The differential on $R_\text{Cartan}$ is
\be
\mathcal D_\text{Cartan}\equiv Q + u^a \hat Q_a
\ee
and it can be checked that there exists a derivation $(s_1)_\text{Cartan}$ on $R$ so that $\mathcal D^2_\text{Cartan} = - [\delta, (s_1)_\text{Cartan}]$ and the cohomology modulo $\delta$ is well-defined\footnote{At this point one could construct a solution to the master equation on $R_\text{Cartan}$ which would look like $\delta + \mathcal D_\text{Cartan} + (s_1)_\text{Cartan}+\cdots$ and thereby make contact with the formulation of \cite{Bonechi:2019dqk,cattaneo2025equivariant}, perhaps by passing to the smaller ring where $\mathcal O_0$ are strictly independent of $\kappa^a$ and invariant.}.

Therefore, semiclassical \textbf{equivariant BV observables}  may be equivalently defined as those which are $ s$-cohomologous to $\mathcal O\in R_\text{Cartan}$ such that $\mathcal O_0$ defines a $\mathcal D_\text{Cartan}$-cohomology class. From this version of the definition it is manifest that equivariant BV observables capture the equivariant  cohomology (with respect to $G$ integrating $\frak g$) of the space of solutions to the equations of motion --- the ``histories''.

\subsubsection{On localisation}
Here I will exhibit a general localisation argument for theories with $\mathcal D^2=0$ off-shell, and outline why there does not appear to exist a model-independent argument in the case where $\mathcal D^2=0 $ on-shell (modulo $\delta$). More specifically I will attempt to localise theories which admit  fermion fields $\hat \psi$ of degree $\rdeg =-1$, so that $\tgh \hat \psi =-1$, and discuss localisation to the locus $Q\hat \psi=0$.

Following the logic of gauge-fixing in the BV formalism, I introduce two trivial pairs and their antifields, with the following degree assignments:
\be
\begin{array}{c|cccc}
                & \xi & \hat\xi & v &  \hat v \\
   \hline \tgh  & 0   & -1      & 1  &  0
\end{array}
\ee
The fields $\hat \xi$ and $\hat v$ are each analogous to antighost fields $\bar c$ and $b$, respectively, in the context of gauge fixing. The localising functional will be
\be
\label{LocFunc}
\langle Q\hat\psi,Q\hat\psi\rangle\equiv (Q\hat\psi)^\dagger Q\hat\psi= \overline{(Q\hat\psi)}^{\bar \alpha} \delta_{\bar \alpha\beta} (Q\hat\psi)^\beta
\ee
where $\langle\bullet,\bullet\rangle$ is a nondegenerate positive definite form. In general this will need to be sesquilinear, hence the dagger $(\dagger)$ notation. (The fact $Q(Q\psi)^\dagger\neq 0$ even if $Q^2=0$ is important. This is hidden in the $\langle Q\hat\psi,Q\psi\rangle$ notation.)

Now take a solution $S_\text{BV}= S_0 +\dots$ to the master equation obtained via homological perturbation as above. The trivial pairs are introduced as follows
\be
S_\text{BV}\to S_\text{BV} + \starred\xi_{\bar \alpha} v^{\bar \alpha} - \starred{\hat \xi}_\alpha \hat v^\alpha
\ee
so that $(S_\text{BV},\hat \xi^\alpha)= \hat v^\alpha$ and $(S_\text{BV}, \xi^{\bar \alpha})= v^{\bar \alpha}$. Consider then the odd symplectomorphism generated by
\be
\Psi \equiv (Q\hat \psi)^\dagger \hat \xi - \xi \hat \xi + \xi \hat \psi\,.
\ee
Then the action transforms to $e^{(\bullet,\Psi)}S_\text{BV}= S_{\text{BV}} + (S_\text{BV},\Psi) +\cdots$. I calculate 
\be
(S_\text{BV},\Psi)\equiv Q_\text{BV}\Psi= \big(Q_\text{BV}(Q\hat\psi)^\dagger-v\big) \hat \xi + \big((Q\hat \psi)^\dagger-\xi\big)\hat v + v\hat \psi + \xi Q_\text{BV}\hat \psi\,.
\ee
 On the lagrangian $\mathcal L$ where the antifields vanish we have
\be
Q_\text{BV}\hat\psi\equiv(S_\text{BV},\hat\psi)=\delta \hat\psi + (\mathcal D-d_W)\hat\psi +s_1\hat \psi + \cdots = Q\hat\psi -\kappa B\hat\psi + u\hat Q\hat\psi
\ee
by antifield number counting. Therefore $Q_\text{BV} \Psi$ will generate a localising functional \eqref{LocFunc} term after  $\xi,\hat v$ and $\hat \xi,v$ are integrated out pairwise. 

When the expansion of the transformed action $e^{(\bullet,\Psi)}S_\text{BV}$ terminates at first order --- which is realised for all theories where $\mathcal D^2=0$ off-shell ---  localisation may be achieved by rescaling $\Psi\to -t \Psi$ and taking the $t\to +\infty$ limit. However, this is not true in the on-shell case: terms $\big((Q_\text{BV},\Psi),\Psi\big)$ and higher nested antibrackets of this form need not vanish, which leads to terms which are higher order in the parameter $t$ and whose positivity is not controlled. One way to proceed is by making assumptions on the expansion of $S_\text{BV}$ in antifields, much like in the recent work \cite{Lysov:2024lge}.  In practice, however, it is easier to localise specific examples using ad-hoc choices of convenient lagrangian submanifolds.

\subsection{Gauge symmetries and equivariant topological twists of type BT}

Let us now consider a theory with a supercharge $Q$ and a bosonic infinitesimal symmetry generator $L$ that obey $Q^2=L$ up to equations of motion. I will call this kind of topological twist a ``BT type'' equivariant twist after Blau and Thompson because they identified this type of equivariant topological supersymmetry as distinct from the DM type in their work \cite{Blau:1996bx} and also studied it earlier \cite{Blau:1991bn}.

\subsubsection{Topological twists and gauge symmetry}
For reasons that will be clear shortly, it is necessary to consider first a (plain) topological twist of a theory with gauge symmetries first. This is of course the obvious generalisation of the discussion of section \ref{sec:topologicaltwist}. We will have three differentials:
\begin{itemize}
    \item $\delta$, the Koszul-Tate differential;
    \item $\gamma$, the vertical differential along the gauge orbits (so that e.g.~for a $\mathrm{U}(1)$ gauge theory, $\gamma A_\mu=\partial_\mu c$ for $c$ the ghost);
    \item and $Q$, the topological supersymmetry.
\end{itemize}
As was explained by Fisch and Henneaux \cite{Fisch:1989rp} the classical BV master action for a theory with gauge symmetry is constructed via homological perturbation of $\delta$ by $\gamma$, for which it is necessary that $\gamma^2=-[\delta, p]$ for some degree-appropriate differential $p$. In the presence of the topological supercharge $Q$ we simply perturb $\delta$ by $\mathcal D\equiv \gamma+ Q$ and run the homological perturbation machine again as in section \ref{sec:topologicaltwist}. To parse the algebraic conditions involved, however, let us establish the gradings carefully:
\be
\begin{split}
    \anti \delta =-1\,,\quad \pdeg \delta= \rdeg \delta =0\,,\\
    \anti \gamma = \rdeg \gamma=0\,,\quad \pdeg \gamma= +1\,,\\
    \anti Q = \pdeg Q=0\,,\quad \rdeg Q= +1\,,
\end{split}
\ee
As is conventional in the BV formalism, ghost number $\gh$ is now refined to $\gh= \pdeg -\anti$, the difference between \emph{pureghost number} $\pdeg$ and antifield number $\anti$. In the absence of a twist, the BV master action has (total) ghost degree $\gh=0$. The BV manifold $M$ where these vector fields act is now the space of fields, antifields, ghosts, and ghost antifields, and we assume that its ring of functions $\cin$ and the symplectic form is trigraded accordingly.

Degree counting shows that $\mathcal D^2=-[\delta,s_1]$ for appropriate  $s_1$ is equivalent to
\be
Q^2=-[\delta,\alpha_{2,0}]\,,\qquad \gamma^2=-[\delta ,\alpha_{0,2}]\,,\qquad [Q,\gamma]=-[\delta, \alpha_{1,1}]
\ee
where I parameterised $s_1= \alpha_{1,0}+ \alpha_{0,1} + \alpha_{1,1}$ with derivations of definite $(\rdeg,\pdeg)$ grading. This tells us that we may perform a topological twist of a gauge theory whenever the supersymmetry involved respects the gauge symmetry (up to equation of motion terms). Homological perturbation then leads to a classical BV master action which is degree zero in the grading $\tgh\equiv \pdeg +\rdeg -\anti$.

\subsubsection{BT-type twists}
\label{sec:BTtwist}
It is in fact not obvious how to deal with the case of a theory with $Q^2=L$ in general. (The obvious differential $Q - \xi L + \frac{\pd}{\pd\xi}$ for odd constant $\xi$ leads nowhere because $H^\bullet(\pd/\pd \xi)=0$.) The cases which Blau and Thompson considered have a special feature, however: $L$ is not arbitrary but it is instead exact in  $\gamma$. To express this invoke a derivation $W$ with the degree assignments:
\be
\begin{array}{c|cccc}
               & \delta & \gamma & Q & W  \\
 \hline \rdeg & 0 & 0 & 1 & 2  \\
 \hline \anti & -1 & 0 & 0 & 0 \\
  \hline \pdeg & 0 & 1 & 0 & -1 
\end{array}
\ee
Now we consider $\mathcal D=\gamma + Q + W$
and calculate $\mathcal D^2=0$, yielding the conditions
\be
Q^2 +[\gamma,W]=0\,,\quad \gamma^2=W^2=[Q,\gamma]=[Q,W]=0\,.
\ee
which respectively encode the vanishing of the $(\rdeg,\pdeg)=(2,0)$, $(0,2)$, $(4,-2)$, $(1,1)$, $(3,-1)$ components of $\mathcal D^2$. This may be completed to a nilpotent differential $s= \delta + \mathcal D +\cdots$ when there exists $s_1$ with $D^2=-[\delta,s_1]$ of appropriate degrees so the conditions just listed only hold on-shell.

Given $Q$ and $W$ are hamiltonian, the homological perturbation lemma constructs $s= \delta + \gamma + Q + W +\cdots$ which is nilpotent and so its hamiltonian $S$ satisfies the usual classical master equation $(S,S)=0$. Thus the appropriate quantum master equation for BT-type equivariantly twisted theories is  the usual one.

\subsection{On BT versus DM}
\label{sec:BTvsDM}

The types defined above are not mutually exclusive: for instance, consider
\be
\mathcal D\equiv d_W+ Q-\kappa^a B_a + u^a \hat Q_a +\gamma + W
\ee
for derivations $d_W,Q,B_a,\hat Q_a,\gamma, W$ as in previous subsections; $\mathcal D^2=0 \mod \delta$ then leads to $Q^2 + [\gamma,W]=0 \mod \delta$ so that we are in the situation of a DM-type equivariant topological twist involving the action of a Lie algebra $\frak g$ with generators $\{T_a\}$ but where the supercharge $Q$ is nilpotent modulo both equations of motion and gauge transformations. The details are easy to work out and lead to formulas like those listed in \eqref{WeilAlgebraOnshell_alternativesigns}.

Moreover we can transmute a DM-type equivariant theory to a BT-type equivariant theory, at the price of treating the generators $\kappa^a,u^a$ of the Weil algebra $W(\frak g)$ as fields to be varied in the path integral. This requires the introduction of antifields $\starred \kappa_a,\starred u_a$ of $\tgh=(-2,-3)$ respectively. Then the Weil differential $d_W$ lifts to a hamiltonian vector field and so does the Kalkman equivariant differential \eqref{KalkmanD}. Writing $Q'$ for the hamiltonian vector field arising thus from the Kalkman equivariant differential, we can define a twist based on the operator
\be
\mathcal D\equiv \gamma + Q' + W
\ee
which is exactly the situation considered for the BT twist. Moreover, the equivariant DM-type classical master equation \eqref{eCME} is easily seen to be equivalent to the ordinary classical master equation for the BT-type theory defined this way.

This last trick is useful whenever we want to consider equivariance with respect to gauge transformations, in which case the equivariant parameters $\kappa^a,u^a$ already appear in the field spectrum even before the twist. We shall see examples of this later. Note that the resulting BT-type theory is not obviously equivalent to the DM-type theory: for instance, expectation values of observables are defined rather differently (see Def.~\eqref{defIntegral}), and the Koszul differential in the resulting BT-type theory also contains information on the equations of motion of $\kappa^a$ and $u^a$.

\section{Two supercharges}
Consider now  the situation of a theory with a doublet of supercharges $Q_A$, $A=1,2$. Topological theories admitting such were called `balanced' TFT's by Dijkgraaf and Moore in the paper \cite{Dijkgraaf:1996tz}, where they also discussed what should be called equivariant balanced topological field theories. Everything in the sequel may be generalised to more than two topological supercharges.

\subsection{Topological twists}
\label{sec:Neq2TopTwists}
The algebra of the supercharges we shall consider will take the form
\be
\label{eq:balancedNONequivcommutator}
[Q_A,Q_B]=-[\delta,\alpha_{AB}]\,,
\ee
for derivations  $\alpha_{AB}=\alpha_{BA}$ of appropriate grading. The bracket is graded and thus  $Q_{1,2}$ are both nilpotent and anticommute  up to equation-of-motion terms.

One can clearly apply the construction of the previous section to realise a topological twist with respect to $Q_1$, $Q_2$, or even a linear combination $v^A Q_A$ (for $v^A\in \mathbb R^2$). But this leads to a family of solutions to the master equation. To define a single theory twisted with respect to both supercharges at the same time we may demand that these solutions to the master equation are homotopic to each other for any $v\neq 0$ in the sense of equation \eqref{eq:homotopyQME}. In practice it seems to suffice to consider the special case where there exists a group action of $\mathrm{SL}(2;\mathbb R)$\footnote{$\mathrm{GL}(2)$ would not make a difference here. Physical considerations, namely of the R-symmetry group of the Euclidean(ised) QFT, however, point towards $\mathrm{SL}(2)$.} on $\mathcal M$ by symplectomorphisms $\varphi_g$ such that $\mathcal D_v\equiv v^A Q_A$ is equivariant, while $\delta$ is invariant:
\be
\label{eq:NiceSL2}
\varphi_{g^{-1}}^\star \mathcal D_v \varphi_{g}^\star = D_{g\cdot v}\,, \quad \varphi_{g^{-1}}^\star \mathcal \delta \varphi_{g}^\star = \delta\,.
\ee
Assuming $[\delta,\mathcal D_v]=0$ and $\mathcal D_v^2=-[\delta , s_1]$ for any specific $v$ and some appropriate $s_1$, which will be the case when \eqref{eq:balancedNONequivcommutator} holds, we obtain an $\mathrm{SL}(2)$ family of homotopic solutions to the master equation. These assumptions entail that the classical action $S_0$ is $\mathrm{SL}(2)$-invariant and that the supercharges $Q_{1,2}$ form an $\mathrm{SL}(2)$ doublet.

The observables of such a theory should be defined to be invariant under $\mathrm{SL}(2)$ up to homotopy so that they are simultaneously annihilated by both supercharges. Such observables are then common to all theories in the family. More explicitly, in the special case discussed above, a classical observable of the theory with $s_v\equiv \delta +\mathcal D_v +\cdots$ should be a BV observable $\mathcal O$ whose antifield independent part $\mathcal O_0$ satisfies
\be
\varphi_g^\star \mathcal O_0=\mathcal O_0 + \mathcal D_v z_g + \delta y_g
\ee
for any $g\in \mathrm{SL}(2)$ and some $z_g, y_g$ depending on $\mathcal O_0$ and $g$. Then the condition that $\mathcal O_0$ is $\mathcal D_v$-invariant mod $\delta$, $\mathcal D_v\mathcal O_0=\delta x$, implies that $\varphi_g^\star \mathcal O_0$ is $\mathcal D_{g\cdot v}$ invariant mod $\delta$. Quantum observables can be defined the same way except we must demand that the BV laplacian $\Delta$ annihilates the hamiltonian for the infinitesimal $\mathrm{SL}(2)$ action.

\paragraph{Remarks.}
\begin{itemize}
    \item The above discussion in fact applies to any number of topological supercharges and any group that acts on them, including discrete groups with some obvious modifications. (Discrete groups will be relevant for the case of equivariant BT-type topological twists.)
    \item Another situation where observables are simultaneously annihilated by all supercharges in the family is where $Q_1$ and $Q_2$ have a bigrading associated with each of them such that $Q_1$ is of bidegree $(1,0)$ and $Q_2$ of bidegree $(0,2)$ which is also respected by the underlying BV manifold $\mathcal M$. Then if we have an $\mathcal O_0$ which has definite bidegree $(m,n)$,
\be
(Q_1+t Q_2)\mathcal O_0=0\iff  Q_1O_0=Q_2O_0=0\,.
\ee
for $t$ any nonzero constant. 
\end{itemize}
\subsection{Equivariant topological twists of type DM}

\paragraph{A ``gormy'' $N=2$ Weil algebra}
Weil algebras for extended supersymmetry have been proposed in \cite{Dijkgraaf:1996tz,Dubois-Violette:1994nxh,Zucchini:1998rz}. I propose an alternative definition that is simpler and has a more transparent geometric interpretation in terms of the geometry of higher differential forms (or ``gorms'').

I start with the observation that the existence of an odd vector field $Q$ which squares to zero on some supermanifold $\mathcal M$ is equivalent to the action of the odd line $\mathbb R[-1]$ seen as the Lie group of odd translations in one variable. Indeed: form $\exp(\theta Q)$ for $\theta$ a formal odd variable of degree $-1$. By the action $\mathcal M\times \mathbb R[-1]\to \mathcal M$ I mean the action of the pullback on the rings of functions given explicitly as follows
\be
\exp(\theta Q)^\star: C^\infty\to C^\infty\otimes \Lambda^\bullet(\mathbb R)\,,\qquad \exp(\theta Q)^\star F=F +\theta Q F
\ee
where $C^\infty\equiv C^\infty(\mathcal M)$ as before and $C^\infty(\mathbb R[-1])$ is the exterior algebra on one variable $\Lambda^\bullet(\mathbb R)$. If $Q$ preserves a grading (e.g.~$\rdeg$ above) on $\mathcal M$ then we have the action of the bigger group $\mathrm{Diff}(\mathbb R[-1])$ --- of transformations $\theta\to \lambda \theta +\eta$ for $\lambda\in\mathbb R$ and $\eta$ odd --- whose bosonic part acts by scaling a function of degree $x$ by $\lambda^x$.

Therefore the desired extended Weil algebra should admit at the very least an action of $(\mathrm{Diff}(\mathbb R[-1]))^2$ and/or $\mathrm{SL}(2)$. I thus define the $N=2$ Weil algebra $W_2(\frak g)$ as
\be
W_2(\frak g)\equiv C^\infty(\operatorname{maps}(\mathbb R[-1]^2,\frak g[1]))
\ee
(and the $N$-extended Weil algebra as $\operatorname{maps}(\mathbb R[-1]^N,\frak g[1])$)\footnote{[\underline{Technical note}: $\operatorname{maps}(M,N)$ refers to the generalised graded/supermanifold in the functor of points approach whose body is the maps of locally superringed spaces $M,N$, roughly speaking. In particular its space of global sections contains the  (graded) ring morphisms $\varphi^\star: C^\infty(N)\to C^\infty(M)$ which in the smooth case, as opposed to the super case, are precisely the pullbacks by smooth functions $\varphi: M\to N$. $\operatorname{maps}$ is the enhancement of $\operatorname{hom}$ that is necessary such that $\operatorname{maps}(\text{point},M)=M$ for any graded/super $M$, which fails for $\operatorname{hom}$ when e.g. $M=\mathbb R[1]$.]}. This is the ring of differential gorms over the shifted Lie algebra $\frak g[1]$ in the terminology of Kochan and \v{S}evera \cite{kochan2003differential,severa2006differential}. It is a direct generalisation of the usual Weil algebra $W_1(\frak g)=W(\frak g)$ which admits the canonical $\mathrm{Diff}(\mathbb R[-1])$ action given by $Q\kappa^a=u^a\,,\, Qu^a=0$. 

The space $\operatorname{maps}(\mathbb R[-1]^2,\frak g[1])$ is a nice finite-dimensional graded supermanifold which is isomorphic to $\frak g[1]\oplus \frak g[3]\oplus \frak g[2]^2$. Upon writing $c^a$ for a linear element of $C^\infty(\frak g[1])\cong \Lambda^\bullet g^\star[1]$, one can abbreviate it as $c\equiv c^a T_a$ where $T_a$ are a basis of $\frak g$. Then an element of $\operatorname{maps}$ is the superfield
\be
c(\theta^1,\theta^2)\equiv \kappa + \theta^A u_A + \theta^{(2)}\xi
\ee
where $\deg \kappa, u_A, \xi= 1,2,3$ and I defined 
\be
\theta^{(2)}\equiv \frac{1}{2}\varepsilon_{AB}\theta^A\theta^B\,.
\ee
Note that $\varepsilon_{AB}$ is the $\mathrm{SL}(2;\mathbb R)$-invariant skew form with $\varepsilon_{12}=1$ and that the degrees here are correlated with supermanifold parity. The upshot is that the $N=2$ Weil algebra is the dgca that is generated by odd generators $\kappa^a$ and $\xi^a$ in degrees 1 and 3 respectively, and by the doublet of even generators $u_A^a$ in degree 2.

This space --- and thus also its ring of functions which is $W_2(\frak g)$ --- clearly has an action of the diffeomorphisms of the odd plane $\mathrm{Diff}(\mathbb R[-1]^2)$ which of course contains $(\mathrm{Diff}(\mathbb R[-1]))^2$. The action of $\mathrm{Diff}(\mathbb R[-1]^2)$ has a degree preserving part which is the obvious $\mathrm{GL}(2;\mathbb R)$ induced from $\theta^A \to M^A_B \theta^B$ for $M\in \mathrm{GL}(2;\mathbb R)$. It reads
\be
\label{eq:gl2}
M^\star \kappa =\kappa\,,\qquad M^\star u_A= M_A^B u_B\,,\qquad M^\star \xi = (\det M) \xi\,.
\ee
There are furthermore two de Rham differentials $\mathrm d_A,\, A=1,2$ induced from the two odd translations $\theta^A\to \theta^A +\eta^A$ and which act as
\be
\label{deRhamMapsg1OddPlane}
\dr_A \kappa=u_A\,,\qquad \dr_A u_B=\varepsilon_{AB}\xi\,,\qquad \dr_A \xi =0\,.
\ee
Finally there are a further two odd generators 
$R_A\equiv u_A^a \pd/\pd \xi^a$.

The two differentials $\mathrm{d}_A$ anticommute and their cohomologies are all trivial except in degree 0 where they equal $\mathbb R$. In fact one can consider them as differentials with respect to a bigrading of $W_2(\frak g)$ where $\dr_1$ has bidegree $(1,0)$ and $\dr_2$ has bidegree $(0,1)$; they then define the desired $(\mathrm{Diff}(\mathbb R[-1]))^2$ action.

\paragraph{On Cartan magic and the lift of the Chevalley-Eilenberg differential.}
The differentials $\dr_A$ seem to differ from the differential $d_W$ of the $N=1$ Weil algebra by terms involving the Chevalley-Eilenberg differential on $\frak g[1]$. In fact the $N=1$ Weil algebra differential can also be presented this way: the redefinition $u\to u +\tfrac{1}{2}[\kappa,\kappa]$ will send the $d_W$ defined above to the simpler differential $Q\kappa^a=u^a,Q u^a=0$. The current presentation will be more convenient.

One can understand the $N=2$ Cartan calculus operations to some extent without invoking the specific form of the differentials on $W_2(\frak g)$ explicitly. What we expect from Moore and Dijgraaf \cite{Dijkgraaf:1996tz} is two de Rham operators along with a ``Lie derivative'' and three contraction operators for each Lie algebra generator $T_a$. 

I will list the complete list of identities that these operators must satisfy later, but I will outline here why we get this bizarre collection of operators for each generator $T_a\in \frak g$. In general if $\varphi\in\operatorname{maps}(\mathcal N,\mathcal M)$ and $X$ is a vector field on the target $\mathcal M$ which has coordinates $c^a$ we may get the explicit expression for the lift $X^\#$ of $X$ to $\operatorname{maps}$ in coordinates from the equality
\be
\label{liftformulaMapping}
\varphi^\star X (c^a)=X^\#(\varphi^\star c^a)
\ee
(The $\varphi^\star$ are usually kept implicit in the superfield notation.)
We have that $X^\#$ is the Lie derivative-analogue (it is indeed the Lie derivative for $\mathcal N=\mathbb R[-1]$). Since $X^\#$ is a vector field on mapping space it may also be multiplied by functions on $\mathcal N=\mathbb R^2[-1]$ which is responsible for the three contraction operators  $\theta^1 X^\#,\theta^2 X^\#,$ and $\theta^{(2)} X^\#$.

\paragraph{The $N=2$ Kalkman equivariant differential.} To arrive at the $N=2$ analogue of the $N=1$ Kalkman equivariant differential $\mathcal D$ \eqref{KalkmanD} it is convenient to split it into mutually anticommuting parts which denote different $\mathrm{Diff}(\mathbb R[1])$ actions. First let us split the Weil differential $d_W$ on $W_1(\frak g)$ into a piece $\dr_1$ that is analogous to the $N=2$ differentials $\dr_A$ given above, and the piece $\dr_\mathrm{CE}^\#$ that lifts the Chevalley-Eilenberg differential:
\be
\label{dWsplit}
d_W= \dr_1 + \dr_\mathrm{CE}^\#
\ee
where I define $\dr_1 \kappa^a =u^a, \dr_1 u^a=0$ and the lift $\dr_\mathrm{CE}^\#$ is the canonical lift of $\dr_\mathrm{CE}$ on $\frak g[1]$ to the mapping space $\maps(\mathbb R[-1],\frak g[1])$, whose ring of functions is $W(\frak g)$, via formula \eqref{liftformulaMapping}. Equivalently, $\dr^{\#}_\mathrm{CE}$ is the CE differential for the canonical Lie algebra structure on the semidirect sum $\frak g\ltimes \frak g[1]$.\footnote{This has the nonvanishing commutation relations $[T_a,T_b]=f_{ab}{}^c T_c$ (for the  $\frak g$ generators) $[T_a,\hat T_b]=  f_{ab}{}^c \hat T_c$ where $\hat T_a$ is in degree $-1$. Degree counting: in my conventions $\frak g[1]$ as a graded manifold has a structure sheaf generated in degree $+1$ only (with generator $\kappa^a$) hence as a graded vector space it must be generated by $\hat T_a$ of degree $-1$.}

We then split the Kalkman equivariant differential \eqref{KalkmanD} into  $V$ and $U$ pieces:
\be
\mathcal D= \underbrace{\dr_{1} + Q}_{\equiv V} + \underbrace{\dr^\#_\text{CE} -\kappa^a B_a +u^a \hat Q_a}_{\equiv U}\,.
\ee
The calculation $\mathcal D^2=0$ is in fact equivalent to $V^2=U^2=\{ V, U\}=0$ so that each of $V$ and $U$ define homological vector fields. They have different interpretations:
\begin{itemize}
    \item $V$ defines a diagonal $\operatorname{Diff}\mathbb R[-1]$ action on the Weil algebra and $C^\infty(\mathcal M)$, simultaneously;
    \item $U$ is the CE differential for the action Lie algebroid of $\frak g\ltimes \frak g[1]$ acting on $\mathcal M$.
\end{itemize}
One proceeds by generalising each of $V$ and $U$ to $N=2$.

The derivation $V$ is easy to generalise: one simply has a pair of operators $\dr_A + Q_A$ where $\dr_A$ acts on $W_2(\frak g)$ and $Q_A$ acts on $\mathcal M$. 

Let us turn to $U$. We need the CE differential for the action Lie algebroid of $(T[1])^2\frak g=\frak g^{\mathbb R^2[-1]}$ and one component of this is the lift of the CE differential of $\frak g$ (on $\frak g[1]$) to a differential $\dr^\#_\text{CE}$ on $\frak g[1]^{\mathbb R^2[-1]}$. (I have switched from the notation $\maps(\mathcal N,\mathcal M)$ to $\mathcal M^{\mathcal N}$ for brevity.) From  formula \eqref{liftformulaMapping} I find
\be
\begin{aligned}
    \dr^\#_\text{CE}\kappa^a = \frac{1}{2}\kappa^b \kappa^c f_{bc}{}^a \,,\qquad \dr^\#_\text{CE}u_A^a=- u_A^b \kappa^c f_{bc}{}^a \,,\\
    \dr^\#_\text{CE} \xi^a= \xi^b \kappa^c f_{bc}{}^a + \frac{1}{2} \varepsilon^{AB} u_A^b u_B^c f_{bc}{}^a\,.
\end{aligned}
\ee

We may then define the action Lie algebroid for $T[1]^2\frak g$ via the differential 
\be
U\equiv \dr^\#_\text{CE} -\kappa^a B_a+ u_A^a \hat Q^A_a + C_a \xi^a
\ee
where $B_a,\hat Q^A_a,C_a$ are all derivations on $C^\infty(\mathcal M)$.\footnote{A way to deduce this expression without working through the definition is that the action Lie algebroid CE differential is always linear in coordinates of the CE algebra --- in this case that of $T[1]^2\frak g$ --- and in the fundamental vector fields of the Lie algebra action.} I then calculate
\be
\begin{split}
    U^2=\frac{1}{2}\kappa^a \kappa^b ([B_a,B_b]- f_{ab}{}^c B_c) - \kappa^a u_B^b ([B_a,\hat Q^B_b] - f_{ab}{}^c \hat Q^B_c)\\
    +\xi^b \kappa^c( [B_c,C_b] + f_{bc}^a C_a) +\frac{1}{2} u_A^a u_B^b ([\hat Q^A_a,\hat Q^B_b]+ \varepsilon^{AB} f_{ab}{}^c C_c)\\
    +u_A^a \xi^b [\hat Q^A_a,C_b] + \frac{1}{2} \xi^a \xi^b [C_a,C_b]\,.
\end{split}
\ee
From here we may read off most of the $N=2$ extended Cartan algebra operators listed by Moore and Dijkgraaf \cite{Dijkgraaf:1996tz}; specifically the ones not involving the de Rham differentials. For the full algebra we also need the following commutator:
\be
[\dr_A+Q_A, U]=\kappa^a [Q_A,B_a] + u^a_B ([Q_A,\hat Q^B_a]-\delta_A^B B_a) + \xi^a (\varepsilon_{AB} \hat Q^B_a - [Q_A,C_a])\,.
\ee
Therefore the $N=2$ Kalkman differential
\be
\label{eq:Neq2Kalkman}
\mathcal D_v\equiv U +v^A (\mathrm d_A + Q_A)
\ee
contains all Cartan operators, and $\mathcal D_v^2=0 \,\forall v \in\mathbb R^2$ encodes the $N=2$ Cartan magic formulas \cite{Dijkgraaf:1996tz} where $B_a$ is identified with their $\mathcal L_a$, $\hat Q_a^A$ with $\iota_a^A$, and $C_a$ with $I_a$.

\paragraph{The master equation.} Fix some specific $v\in \mathbb R^2$ and assume
\be
\mathcal D_v^2=-[\delta , s_1]\,,\qquad [\delta, \mathcal D_v]=0
\ee
so that the $N=2$ Cartan magic formulas are weakened to only hold up to homotopy and the action $S_0$ is invariant under the operators $Q_A$, $B_a$, $\hat Q_a^A$ and $C_a$. Then homological perturbation yields the nilpotent differential
\be
s_v\equiv \delta + \mathcal D_v + s_1 +\cdots
\ee
whence we deduce the $N=2$ equivariant classical master equation of DM type:
\be
(S_\text{BV},S_\text{BV})+ 2 d_{W} S_\text{BV}=0
\ee
with
\be
\label{eq:dWeilNeq2DirV}
d_W \equiv \mathrm d^\#_{\text{CE}}+ v^A \mathrm{d}_A
\ee
and $S_\text{BV}$ the hamiltonian function of $s_v- d_W$. Similarly, the quantum master equation for a quantum observable $f$ (including $f=\exp(S_\text{BV}/\hbar)$) takes the form
\be
(\hbar \Delta +d_W)f=0
\ee
and leads to an equation for $S_\text{BV}$ identical to \eqref{eQME} but for $d_W$ as in \eqref{eq:dWeilNeq2DirV}.

Note that \textbf{both} $S_\text{BV}$ and $d_W$ depend on $v$ in $\mathbb R^2$. The solutions $S_\text{BV}$ are defined to be pairwise homotopic for any nonzero value of $v$ for reasons discussed in subsection \ref{sec:Neq2TopTwists}. (The remainder of this paragraph explains how this works but is rather dry.) To account for the fact that the parameter $v$ will be evolving under homotopy we must generalise the definition of homotopy from \eqref{eq:homotopyQME} to the two conditions
\be
(\hbar \Delta +D_W+\mathrm d)^2=0\,,\qquad (\hbar \Delta +D_W+\mathrm d)F=0
\ee
where $D_W\equiv d_W + X\dr t$, $X$ is a degree-preserving derivation on $W_2(\frak g)$, $\mathrm d= \frac{\dr}{\dr t}\dr t$, $F=f + g\dr t$, and where $v,X, F$ are allowed to depend on $t\in [0,1]$. The first condition determines the evolution of $d_W=\mathrm{d}^\#_\text{CE} + v^A\mathrm d_A$:
\be
\frac{\dr}{\dr t}v^A \mathrm d_A=[d_W  , X]\,.
\ee
This equation is easily solved via the canonical $\mathrm{SL}(2)\into \mathrm{GL}(2)$ action on the differentials $\mathrm{d}_A$ induced by the ``gormy'' $\mathrm{GL}(2)$ of eq.~\eqref{eq:gl2}. In the nice scenario described around eq.~\eqref{eq:NiceSL2}, we must therefore combine the $\mathrm SL(2)$ acting by symplectomorphisms $\varphi_g$ on $\mathcal M$ with the one that acts on $\operatorname{maps}(\mathbb R[-1]\times \mathbb R[-1],\mathfrak g[1])$ by permuting the doublet of odd coordinates $\theta^A$. In this nice scenario, solutions to the equivariant master equation for any value of $v$ are homotopic to the solution for any \emph{fixed} value of $v$\footnote{Because rotations by $2\pi n$ in $\mathrm{U}(1)$ within $\mathrm{SL}(2)$ act trivially on $\mathbb R^2$; therefore even though $\mathbb R^2-\{0\}$ is not simply connected there can be no holonomy when going around the origin.}, which is a little stronger than them being pairwise homotopic.

\paragraph{Integration.}
This may be defined exactly as in Definition \ref{defIntegral} and enjoys the same properties that integral does. A crucial point is that $d_W$ acting on $W_2(\frak g)$ is acyclic for any value of $v\in \mathbb R^2$ and $H^0(d_W)\cong \mathbb R$ so that expectation values are insensitive to which $v\in \mathbb R^2$ is being used.  (The only thing that changes for different values of $v$ is the identification of trivial pairs in $W_2(\frak g)$.) This is a key difference from the formulation of Dijkgraaf and Moore (where the analogues of $\mathrm d_{1,2}$ are not acyclic; see section 3.5 of \cite{Dijkgraaf:1996tz}) and motivates the ``gormy'' definition of $W_2(\frak g)$ I gave earlier.

\subsection{Equivariant topological twists of type BT}
\label{sec:Neq2BT}
Compared to type-DM equivariant twists these are much less convoluted to formulate. An $N=2$ type-BT equivariant twist can be defined whenever there exists an odd derivation $\mathcal D_v$ of $\rdeg=1$ of the form
\be
\label{eq:DvNeq2BT}
\mathcal D_v= \gamma + v^A Q_A +\frac{1}{2} v^A v^B \delta_{AB} W
\ee
such that $\mathcal D_v^2=-[\delta,s_1]$ for all $v\in\mathbb R^2$, as usual.  This means that the supersymmetries $Q_{1,2}$, vertical differential $\gamma$ along the gauge orbits, and  odd derivation $W$, all with the degree assignments of section \ref{sec:BTtwist}, must satisfy the identities
\be
\gamma^2=W^2=[Q_A,\gamma]=[Q_A,W]=0\,,\qquad [Q_A,Q_B]=-\delta_{AB} [\gamma,W]
\ee
modulo $\delta$. When $[\delta, \mathcal D_v]=0$ as well, we obtain solutions to the  classical master equation for each value of $v$.

The above ansatz for $\mathcal D_v$ could be generalised by, say, replacing $ v^A v^B \delta_{AB} W\to W_{AB}$. This does not seem necessary in order to accommodate known examples of such theories, however.

\section{Examples}
\subsection{Supersymmetric quantum mechanics}
Let us consider a SUSY quantum mechanics with superpotential $h(x)$ and action \be
\label{S0SQM}
S=S_0[x,\phi,\hat\psi]=\int \dr t\; \Big( \frac{1}{2} \dot x^2 +\hat \psi \dot \psi +\frac{1}{2} h'{}^2 - h''\hat\psi \psi\Big)\,.
\ee
The expression $x=x(t)$ is real-valued while $\psi$ and $\hat \psi$ are Grassmann-odd. The ghost number grading reads $\operatorname{gh}x=\operatorname{gh}\psi=\operatorname{gh}\hat\psi=0$. There is no gauge-symmetry so $S$ is also the BV master action. The antibrackets are defined as usual
\be
(x(t),\starred x(t'))=(\psi(t),\starred\psi (t'))=(\hat \psi(t),\starred{\hat \psi}(t'))=\delta(t-t')
\ee

This action has the  odd left symmetry $Qx=\psi, Q\psi=0,Q\hat\psi=-\dot x + h'$ with 
\be
Q^2 \hat \psi =-\dot \psi +h'' \psi\,,\qquad Q^2 x=Q^2 \psi=0\,,
\ee
where we recognise the equation of motion for $\psi$. 

\paragraph{The topological twist.} We have $\Theta=-\int dt\; \psi \starred x + (-\dot x + h') \starred{\hat \psi}$ as the hamiltonian of $Q$, and its antibracket is
\be
(\Theta,\Theta)=\int \dr t\; 2(\dot \psi - h''\psi)\hat \psi^\star = \Big(S_0,\int\dr t\; -\hat\psi^\star{}^2 \Big)\,.
\ee
(Boundary terms were dropped. I justify this later.)
This calculation implies $Q^2$ is $\delta$-exact. Moreover there exists a grading $\rdeg$ with $\operatorname{Rdeg} \psi =1, \operatorname{Rdeg} x=0, \operatorname{Rdeg} \hat \psi =-1$, and so on, so that $\operatorname{Rdeg} Q=+1$. Therefore $\delta=(S_0,\bullet)$ and $Q$ satisfy the assumptions of the homological perturbation lemma and we are in the situation of section \ref{sec:topologicaltwist}. Following the algorithm described therein one calculates the classical BV master action for the topologically twisted theory (I again write $S_\text{BV}$ instead of $S_\text{tBV}$)
\be
S_\text{BV}= S_0 +\Theta +\int \dr t\; \frac{1}{2} \starred{\hat \psi}^2\,.
\ee
This is of degree zero in the twisted ghost number grading $\tgh=\gh + \rdeg$. Note that the various gradings are as follows
\be
\begin{array}{c|cccccc} & x & \starred{x} & \psi & \starred \psi & \hat\psi & \starred{\hat\psi} \\
    \hline \gh & 0 & -1 & 0 & -1 & 0 & -1
\\ \hline \rdeg & 0 & 0 & 1 & -1 & -1 & +1
\\ \hline \tgh  & 0 & -1 & 1 & -2 & -1 & 0 \end{array}
\ee
The Grassmann parity is correlated with $\tgh \mod 2$, so that the BV master action for the twisted theory has the degree assignments of the BV master action of a purely bosonic theory. Note however that $\starred{\hat \psi}$ is not real-valued but is instead a generator of the ring $\mathbb R[\starred{\hat \psi}]$ of formal power series.

By the observation about observables in section \ref{sec:topologicaltwist}, the observables of this BV theory are isomorphic to the $\gh=0$ functionals that are $Q$-invariant modulo $\delta$, i.e.~modulo terms that vanish by equations of motion. This gives $\psi=0$ and $\dot x= h'$ (which imply the equation of motion for $x$, namely $\ddot x=h'h''-h^{(3)} \hat \psi \psi$ which, ignoring the fermions, is equivalent to $\dr ( \dot x^2-h'{}^2)/\dr t=0$ --- choosing the other supersymmetry localises to the other ``half'' of the equation of motion.)

Note that $S_\text{BV}$ also trivially solves the quantum BV master equation, since $\Delta S_\text{BV}=0$. (Here $\Delta$ is the naive expression $\int \dr t 
\;{\delta^2}/{\delta \phi(t)\delta \starred\phi(t)}$ which happens to be well-defined acting on this $S_\text{BV}$.)

\paragraph{The equivariant twist.} We can also formulate the (DM-type) equivariantly twisted theory. The relevant infinitesimal symmetries and their hamiltonians are
\begin{subequations}
    \begin{alignat}{2}
        Q&\equiv (\Theta,\bullet)\,,\quad \Theta\equiv -\int dt\; \psi \starred x + (-\dot x + h') \starred{\hat \psi} \\
        \hat Q&\equiv (\hat \Theta,\bullet)\,,\quad \hat \Theta\equiv -\int dt\; (-\hat\psi )\starred x + (\dot x + h') \starred \psi\,,\\
        B&\equiv (H,\bullet\,,)\quad H\equiv -\int dt\;  2\dot x \starred x + 2\dot \psi \starred \psi+ 2 \dot{\hat \psi} \starred{\hat \psi}\,,
    \end{alignat}    
\end{subequations}
Of these $B$ is an infinitesimal time translation, and $\hat Q$ is the other supersymmetry of the particle mechanics model \eqref{S0SQM}.

We need to confirm the identities of equation \eqref{WeilAlgebraOnshell_alternativesigns} (the Cartan calculus identities modulo $\delta$) for these symmetries. Their algebra is obtained by straightforward calculation. We have
\be
(\hat \Theta,\hat \Theta)=  \Big(S_0,\;\int\dr t\; -\psi^\star{}^2 \Big)
\ee
and 
\be
(\Theta,\hat\Theta)= H- \Big(S_0,\int\dr t\;\hat\psi^\star \psi^\star \Big)\,.
\ee
We also need to establish identities \eqref{identityB} and \eqref{identityE} i.e.~we need to establish that $B$ commutes with $Q$ and $\hat Q$ on-shell. Since its hamiltonian $H$ realises the time derivative on fields and antifields both, and since $\Theta$, $\hat \Theta$ are both local in time, this is true if we drop total time derivatives. That manipulation is valid when the time direction is compact\footnote{Admittedly we could instead have something like fields on $\mathbb R$ with compact support, but I do not care for that, do you?} and when \emph{both} bosons $x$ and \emph{fermions} $\psi,\hat \psi$ are \emph{time-periodic} (this is necessary because $\Theta,\hat \Theta$ are linear in fermions and their antifields are bosonic). This means \emph{the equivariant twist of SUSY quantum mechanics ``selects'' the Witten index}.

Putting all the ingredients together, the solution to the equivariant classical master equation \eqref{eCME} is
\be
S_\text{BV}\equiv S_0 + \Theta - \kappa H + u\hat \Theta + H_{s_1}
\ee
where $H_{s_1}=\int\dr t\; \frac{1}{2}(\starred{\hat\psi} + u \starred{\psi})^2$: indeed one calculates
\be
(S_\text{BV}, S_\text{BV})= 2 u H + 2(\Theta + u\hat \Theta,H_{s_1})=2 u H
\ee
(there is a cancellation that is special to the SQM model here) and
\be
d_W S_\text{BV}=-u H\,.
\ee
Here $\kappa,u$ are the generators of the Weil algebra for time translations, which is 1-dimensional.


Finally we note that, again, $\Delta S_\text{BV}=0$ so that this action also solves the equivariant BV master equation \eqref{eQME}. This requires regularising the BV laplacian. The naive BV laplacian is only singular when applied to the generator $H$ of time translations inside $S_\text{BV}$. It is trivial to see that any regularisation of the form
\be
\Delta_\varepsilon\equiv \int \dr t \int \dr t' \; \delta_\varepsilon(t-t') \frac{\delta}{\delta \phi(t)}\frac{\delta}{\delta \starred{\phi}(t')}\,,
\ee
with $\delta_\varepsilon(t-t')$ a nascent Dirac delta function,  annihilates $H$ as well as all the other terms in $S_\text{BV}$. This is true even before the limit $\varepsilon \to 0^+$. This class of regularisations contains the heat-kernel regularisation used by Costello \cite{costello2011renormalization}.

\paragraph{Localisation.} This is a theory which is easily localised by an ad-hoc choice of lagrangian submanifold. The key observation is that the solution to the (equivariant) classical master equation involves a term $H_{s_1}$ quadratic in fermion antifields $\hat \psi$, which can generate a perfect square. I choose the exact lagrangian submanifold associated to the ``gauge-fixing'' fermion
\be
\lambda\Psi\equiv \lambda\int \dr t\; \hat \psi Q\hat \psi= \lambda\int \dr t\;\hat \psi(h'-\dot x)\,.
\ee
This is of degree $-1$ in the twisted grading $\tgh$. The ``gauge-fixed'' action reads
\be
S_\text{g.f.}\equiv S_0 +  \int \dr t\, \lambda(\dot x-h')^2 +  \lambda\hat \psi (\dot \psi - h''\psi)+ \frac{1}{2}\lambda^2
(\dot x-h')^2  +\cdots
\ee
where I have omitted any terms involving $\kappa$ and $u$ because they are not relevant for the localisation argument. Since the path integral is invariant under perturbations of $\lambda$ we may evaluate the path integral of $\exp(-S_\text{g.f.}/\hbar)$ in the $\lambda \to \infty$ limit where the leading Gaussian term 
\be
-\frac{\lambda+\lambda^2}{2}\int\dr t
(\dot x-h')^2= -\frac{\lambda+\lambda^2}{2}\int \dr t\,\Big(\dot x^2 + h'{}^2 + \frac{\dr}{\dr t}(\cdots)\Big)
\ee
dominates. Thus the path integral restricts to configurations with $\dot x= h'=0$.

We have found, therefore, that the path integral localises to the critical points of the superpotential. Upon calculating the 1-loop determinant with more-or-less standard manipulations we find the following result for the path integral $Z$ without any operator insertions:
\be
Z=\sum_{x_0 \text{ critical}}\operatorname{sign} h''(x_0)
\ee
assuming that the superpotential is Morse. This agrees with the Witten index for this SUSY quantum mechanics, which accords with the earlier observation that the equivariantly twisted theory makes sense only for periodic fermions.

\subsection{Topological A- and B-models}
Here I will show how the twists of the supersymmetric non-linear two-dimensional sigma model fit into the topological twist framework I have been advocating for. I will follow the review \cite{Witten:1991zz}. 

The supersymmetric sigma model action with a K\"ahler target space can be written as
\be
\label{eq:S0sigmamodel}
S_0=\int \dr^2z\, \frac{1}{2}g_{i\bar j} (\pd \phi^i \bar\pd \phi^{\bar j} + \bar \pd \phi^i \pd \phi^{\bar j})
 + g_{\bar i j}(\psi_-^{\bar i}D \psi_-^j + \psi_+^{\bar i} \bar D \psi_+^j) - R_{i\bar jk\bar \ell} \psi_+^i \psi_+^{\bar j} \psi_-^k \psi_-^{\bar \ell}\,.
\ee
(Note that I am using a complex conjugation operation for fermions where they do not swap places when conjugated, thereby eliminating factors of $i$. See appendix \ref{appendix:CC}.) The fields are $2 d$ bosonic scalars  written in complex notation as $(\phi^i,\phi^{\bar i})$ ($i=1,\dots d$), and the fermions $\psi_\pm^i$ and $\psi_\pm^{{\bar i}}$. The model enjoys $N=2$ supersymmetry and the fermions transform under the $\mathrm{U}(1)\times \mathrm{U}(1)$ R-symmetry group as
\be
\psi_\pm^i \to \exp\big(i(a_V \pm a_A)\big) \psi_\pm^i\,,\qquad \psi_\pm^{\bar i} \to \exp\big(i(-a_V \mp a_A)\big) \psi_\pm^{\bar i}
\ee
where $a_V\in\mathbb R$ is the transformation parameter for the diagonal (vector) subgroup and $a_A \in\mathbb R$ the transformation parameter for the antidiagonal (axial) subgroup.

The A- and B-twists are realised in the framework of section \ref{sec:topologicaltwist} by identifying the grading $\rdeg$ with the charge of the fermions under the $\mathrm{U}_A(1)$ or $\mathrm{U}_V(1)$ subgroups. Explicitly:
\be
\begin{aligned}
    &\text{A-twist:}&  &\rdeg \psi_+^i=+1\,,& &\rdeg \psi_-^i=-1\,,& \\
    &\text{B-twist:}&  &\rdeg \psi_+^i=-1\,,& &\rdeg \psi_-^i=-1\,,&
\end{aligned}
\ee
and the opposite $\rdeg$ is assigned to the barred fields.
Note that it is the ``oppposite'' $\mathrm{U}(1)$ that appears in $\rdeg$ relative to the $\mathrm{U}(1)$ which appears in the twisting of the Lorentz generator, so that the supersymmetry with $\rdeg=+1$ in the A-twist is the one which turns scalar after twisting the Lorentz generators by the vector symmetry (vice versa for the B-twist). These supersymmetries are given by
\begin{subequations}
    \begin{align}
    Q \phi^i &= \psi_+^i\,,\\
    Q \phi^{\bar i} &= \psi_-^{\bar i}\,,\\
    Q \psi_+^{\bar i} &=-\pd \phi^{\bar i} -\psi_-^{\bar j}\Gamma^{\bar i}{}_{\bar j \bar k} \psi_+^{\bar k} \,,\\
    Q \psi_-^{i} &=-\bar\pd \phi^{i} -\psi_+^{j}\Gamma^{ i}{}_{ j  k} \psi_-^{ k} \,,\\
    Q \psi_+^i&=0\,,\\
    Q \psi_-^{\bar i} &=0
    \end{align}
\end{subequations}
for the A-twist, and
\begin{subequations}
    \label{eq:QBtwist}
    \begin{align}
    Q \phi^i &= 0\,,\\
    Q \phi^{\bar i} &= \psi_+^{\bar i}+\psi_-^{\bar i}\,,\\
    Q \psi_+^{\bar i} &=\psi_+^{\bar j}\psi_-^{\bar k} \Gamma^{\bar i}_{\bar j \bar k}\,,\\
    Q \psi_-^{i} &=-\bar \pd\phi^i \,,\\
    Q \psi_+^i&=-\pd\phi^i\,,\\
    Q \psi_-^{\bar i} &=-\psi_+^{\bar j}\psi_-^{\bar k} \Gamma^{\bar i}_{\bar j \bar k}\,
    \end{align}
\end{subequations}
for the B-twist.

For the B-twist we have $Q^2=0$ off-shell\footnote{To see this, and also to resolve the calculations for the A-model, it is necessary to use the fact the target manifold is K\"ahler and in particular the vanishing of all Riemann tensor components except $R_{i\bar j k \bar\ell}$. For the B-model the key identity is $R_{\bar i \bar j \bar k \ell}=0$.}. For the A-twist the relevant supersymmetry is only nilpotent modulo on-shell terms. The precise relation is
\be
Q^2=-[\delta, s_1]\,,\qquad s_1=\Big(\int d^2 z
\; g^{i\bar j} \starred\psi_{-i}\starred\psi_{+\bar j}\,, \bullet\Big)
\ee
where the antifields $\starred\psi_{-i}$ and $\starred\psi_{+\bar j}$ are the canonical conjugates to $\psi_-^i$ and $\psi_+^{\bar j}$ respectively (and $\delta\equiv (S_0\,,\bullet)$). In fact the expansion in antifield number $s= \delta + Q + s_1+\dots$ terminates at $s_1$ since $s_1^2=[Q,s_1]=0$, $Q$ here being the cotangent lift of the A-twist topological supersymmetry. Therefore the classical BV master action for the A-model is
\be
S_{\text{A-model}}\equiv S_0 + \Theta + \int d^2 z
\; g^{i\bar j} \starred\psi_{-i}\starred\psi_{+\bar j}\,,
\ee
$\Theta$ being the hamiltonian for $Q=(\Theta,\bullet)$.

\paragraph{Localisation.} Since the B-model has an off-shell nilpotent topological supersymmetry, I will only discuss localisation for the A-model. One way to do it is completely analogous to the treatment of SUSY quantum mechanics in a previous subsection: write down the obvious ``gauge-fixing'' fermion $\Psi$ of degree $-1$ in the twisted grading $\tgh=\gh+\rdeg$:
\be
\Psi\equiv \int \dr^2 z\; \frac{1}{2} g_{i\bar j} (\psi_-^i Q\psi_+^{\bar j} + \psi_+^{\bar j} Q \psi_-^i)\,.
\ee
Upon evaluating the exponential of the BV master action on the exact lagrangian submanifold generated by $\lambda \Psi$ ($\lambda \in\mathbb R$) I find the following leading Gaussian term in the $\lambda \to \infty$ limit,
\be
\int \dr^2z\; g_{i\bar j} \bar{\pd}\phi^i \pd \phi^{\bar j}\,,
\ee
from which it follows that that the path integral localises to configurations with $g_{i\bar j} \bar{\pd}\phi^i \pd \phi^{\bar j}=0$. This equation is equivalent to the worldsheet instanton equations
\be
\bar \pd \phi^i = \pd \phi^{\bar i}=0
\ee
describing holomorphic maps.

\subsubsection{An equivariant B-model}
\label{sec:equivariantBmodel}
I will now define a DM-type equivariant version of the B-model which falls under the framework of section \ref{sec:equivariant}. The symmetry with respect to which the new model will be equivariant is the holomorphic $\mathrm{U}(1)$ symmetry
\be
z\to e^{-i\theta}z\,,\qquad \bar z \to e^{i\theta} \bar z
\ee
on the worldsheet ($z$ being a holomorphic coordinate). This group action is globally well defined on the sphere, where its fixed points $z=0$ and $z=\infty$ correspond to the north and south poles via stereographic mapping, but it not defined on the torus or on higher genus worldsheets. For this reason I will assume the worldsheet is now $S^2$.

For the B-twist the fermions are taken to lie in the bundles
\be
\begin{split}
    \psi_+^i \in \Gamma[T^\star_{(1,0)} S^2\otimes \Phi^\star T^{(1,0)} X]\,,\quad \psi_-^i \in \Gamma[T^\star_{(0,1)} S^2 \otimes \Phi^\star T^{(1,0)} X]\\
    \psi^{\bar i}_+,\psi^{\bar i}_- \in \Gamma [\Phi^\star(T^{(0,1)} X)]
\end{split}
\ee
which is to say that the $\psi_\pm^{\bar i}$ are scalars valued in $T^{(0,1)}$ of the target K\"ahler space $X$, $\Phi$ being the map $S^2\to X$ parameterised by $\phi^i,\phi^{\bar i}$, while $\psi_\pm^i$ are the components of the 1-form $\rho^i=\psi_+^i \dr z + \psi_-^i \dr \bar z$ on the worldsheet, valued in $T^{(1,0)}$, as the $i/\bar i$ index indicates. (The point of this bundle assignment being that the supercharge $Q$ is well defined --- as a scalar --- on worldsheets of any genus \cite{Witten:1991zz}.)

With these assignments the infinitesimal $\mathrm{U}(1)$ transformations on the fields read
\be
\begin{split}
    B \psi_\pm^{\bar i}= i \bar z \bar\pd\psi_\pm^{\bar i}-iz \pd \psi_{\pm}^{\bar i}\,,\qquad
    B \psi_\pm^{ i}=i \bar z \bar\pd\psi_\pm^{ i}-iz \pd \psi_{\pm}^{ i} \mp i \psi_\pm^i\,.
\end{split}
\ee
To construct the equivariant B-model is to find another fermionic operator $\hat Q$ such that the Cartan calculus identities \eqref{WeilAlgebraOnshell_alternativesigns} are satisfied (for some $\alpha,\beta,\gamma,\dots$ that parameterise the failure of the identities to hold strictly off-shell) and also $\hat Q$ must be a symmetry of the action $S_0$.

Given the ansatz $\hat Q \phi^{\bar i}=0$, some pen-and-paper struggle determines $\hat Q$: 
\begin{subequations}
    \begin{align}
    \hat Q \phi^i &= i z \psi_+^i - i \bar z \psi_-^i\,,\\
    \hat Q \phi^{\bar i} &= 0  \,,\\
    \hat Q \psi_+^{\bar i} &=-iz\pd\phi^{\bar i}\,,\\
    \hat Q \psi_-^{i} &=-i z \psi_+^j \psi_-^k \Gamma_{jk}^i \,,\\
    \hat Q \psi_+^i&=-i \bar z \psi_+^j \psi_-^k \Gamma_{jk}^i\,,\\
    \hat Q \psi_-^{\bar i} &=+i\bar z\bar\pd \phi^{\bar i}\,.
    \end{align}
\end{subequations}
For this $\hat Q$ and the $Q$ associated to the B-twist \eqref{eq:QBtwist} the independent Cartan calculus identities of equation \eqref{WeilAlgebraOnshell_alternativesigns} are satisfied strictly except for
\be
B=[Q,\hat Q]+ [\delta,\gamma]\,;\qquad \gamma\equiv(\Gamma,\bullet)\,,\quad \Gamma\equiv -i\int \dr^2 z\;  g^{i\bar j}(\starred\psi_{+\bar j} - \starred{\psi}_{-\bar j}) ( z\starred \psi_{-i}+\bar z \starred \psi_{+i})\,.
\ee

Note that $\hat Q$ takes the form of one of the supersymmetries of the sigma model with action $S_0$, see \cite[formula (2.5)]{Witten:1991zz}, specifically one with spinor parameters $\tilde \alpha_\pm=0$ and $\alpha_+=z$, $\alpha_-=-\bar z$. (These are indeed sections of the spinor bundles $\mathcal O(1),\overline{\mathcal O(1)}$ on $S^2=\mathbb CP^1$.) Therefore $\hat Q$ is also an invariance of the action. However its geometric interpretation changes under the B-twist: if $r=i\bar z\bar \pd - i z \pd$ denotes the vector field generating the $\mathrm{U}(1)$ action, $\hat Q \phi^i$ is the contraction $\iota_r \rho^i$ for $\rho^i=\psi_+^i \dr z + \psi_-^i \dr \bar z$, $Q\psi_+^i$ is the $(1,0)$ component of the contraction $\iota_r(\psi_+^j \dr z\wedge \psi_-^k \dr \bar z \Gamma_{jk}^i)$, and so on for the rest of the variations, so that $\hat Q$ is a scalar operator instead of a spinor.

The above discussion suffices to define the equivariant B-model, at least at the level of a solution of the classical equivariant master equation \eqref{eCME}. I will not display this solution explicitly; this, and other models obtained in a similar way, will be studied in the future. I will however make some remarks:
\begin{enumerate}
    \item In spite of the fact $Q^2=\hat Q^2=0$ off-shell, the equivariance is realised up to homotopy of the Koszul differential $\delta$. One could in principle find auxiliary fields such that the equivariance is realised off-shell but it is not clear whether that requires any less guesswork than determining $\hat Q$ given $Q$ and $B$. 

    \item Another proposal for an equivariant B-model was made very recently by Festuccia, Mauch, and Zabzine \cite{Festuccia:2024zoh} based on the works \cite{Yagi:2014toa,Nekrasov:2018pqq}. It employs a topological supercharge they call $\delta$, whose cohomology figures in the determination of observables, and another supercharge $\hat \delta$. The first supercharge seems to be related to the supercharges of this work via $\delta=Q+\hat Q$ (compare the variations of the scalars $\phi^i$ and $\phi^{\bar i}$). It is less clear how to relate $\hat \delta$ to our supercharges, or that the observables defined in section \ref{ObservablesSection} are equivalent to the ones defined in \cite{Festuccia:2024zoh}.
\end{enumerate}

\subsection{Twists of 4-dimensional super Yang-Mills theories}
Here I shall show that twists of SYM theories can fit into the formalism. The point is to exhibit that I have not been theorising about the empty set, so the discussion will be very brief. I will focus on theories admitting two topological supercharges.

\paragraph{The Vafa-Witten twist.} The Vafa-Witten twist of maximal 4-dimensional super Yang-Mills theory was the motivating example for Dijkgraaf and Moore's ``balanced'' topological field theory construction. In terms of my classification it is \emph{almost} an equivariantly DM-type twisted theory with two topological supercharges: the existence of operators $Q_A, B_a, \hat Q^A_a, S_a$ such that the $N=2$ Kalkman equivariant differential of \eqref{eq:Neq2Kalkman} is nilpotent is established in \cite{Dijkgraaf:1996tz} already. Actually since the equivariance is with respect to gauge symmetry, the equivariant parameters $\kappa^a$ and $u^a$ in my notation need to be promoted to fields and so we land at an equivariant BT-type $N=2$ theory as explained in subsection \ref{sec:BTvsDM}. (Of course for the Vafa-Witten theory there is no issue realising the $N=2$ Cartan calculus identities off-shell.)

\paragraph{The GL twist.} Here I will point out the existence of a theory with two topological supercharges for which no auxiliary fields which close the algebra off-shell are known. Such a theory is therefore a great example for the BV twist machinery developed in this paper. The theory is what is known as the Marcus twist of $N=4$ super-Yang-Mills theory \cite{Marcus:1995mq} (although the existence of the twist was pointed out by Yamron \cite{Yamron:1988qc} much earlier) which is also sometimes known as the Geometric Langlands (GL) twist since the influential work of Kapustin and Witten \cite{Kapustin:2006pk}.

The superalgebra is listed in the work \cite{Blau:1996bx}, equations (4.18)-(4.20). In their notation, there are two topological supercharges $Q$ and $\bar Q$, with
\be
Q^2=\bar Q^2= L_\phi\,,\quad [Q,\bar Q]=\int \dr^4 x\; 2i\frac{\delta S}{\delta \chi(x)} \frac{\delta}{\delta \chi(x)} -2i \Big[\frac{\delta S}{\delta B(x)},\phi(x)\Big]\frac{\delta S}{\delta B(x)}\,,
\ee
 $S$ being their action (4.16)  ($S_0$ in my notation), and $L_\phi$ denoting a gauge transformation with parameter $\phi$, such that $ L_\phi A=\dr \phi + [A.\phi]$. The classical fields and their degree assignments such that $S_0$ has $\rdeg =0$ are
\be
\begin{array}{c|ccccccccccc} & A & \psi & \phi & \bar \phi & \eta & V & \bar \psi & \bar \eta & u &\chi & B
\\ \hline \rdeg & 0 & 1 & 2 & -2 & -1 & 0 & 1 & -1 & 0 &-1 & 0
\end{array}
\ee
while their Grassmann parity is $\rdeg\mod 2$. (In terms of the discussion of \cite{Kapustin:2006pk}, $\rdeg$ is the $\mathrm{U}(1)$ eigenvalue of the generator they call $\mathcal K$ in their (3.3).) From the ghost and antifield sectors we will only need the ghost $c$ for the Yang-Mills gauge symmetry of the gauge potential $A$. This is encoded by the vertical differential $\gamma$ that acts in the standard way
\be
\gamma A=\dr c + [A,c]\,,\quad \gamma \bar \phi=-[c,\bar \phi]\,,\quad \cdots
\ee
and has $\rdeg \gamma=0\,,\pdeg\gamma = +1$. Here the brackets $[\bullet,\bullet]$ denote commutators in the Lie algebra $\frak g$ where all fields are valued.

We are in the situation of extended $N=2$ topological equivariant supersymmetry of type BT of subsection \ref{sec:Neq2BT}. To evidence this we first notice the differential $W$ that acts on the space generated by the classical fields and ghosts as
\be
W c=\phi\,,\qquad W(\text{else})=0\,.
\ee
which has the correct degree assignment $\rdeg W=2, \pdeg W=-1$ and is such that $L_\phi=[\gamma, W]$. Moreover we have $[Q,\bar Q]=-[\delta,\alpha]$ for $\alpha$ the differential defined on generators by $\alpha \chi =-2i \starred\phi\,, \alpha B=2i [\starred B,\phi]$, $\alpha(\text{else})=0$. Therefore we may construct a family of solutions to the (classical) master equation parameterised by $ v\in\mathbb R^2$ appearing in the differential $\mathcal D_v$ of eq.~\eqref{eq:DvNeq2BT}, upon identifying $Q_1\equiv Q, Q_2\equiv \bar Q$.

A subtlety in this particular example is that there does not exist a nice $\mathrm{SL}(2)$ symmetry on the space of fields that induces an $\mathrm{SL}(2)$ doublet action on $Q_{1,2}$. There is however a $\mathbb Z_2$ action that swaps $Q$ with $\bar Q$ which does arise from a transformation of the fields, and this is described explicitly in \cite[formula (4.17)]{Blau:1996bx}. This $\mathbb Z_2$ is present on manifolds admitting orientation-reversing symmetries and is the equivalence $t\to t^{-1}$ described by Kapustin and Witten \cite{Kapustin:2006pk}, in terms of their parameterisation $Q + t \bar Q$ of the space of supercharges. This $\mathbb Z_2$ lies in the $\mathrm{O}(2)\into \mathrm{SL}(2)$ which preserves the bilinear form $\delta_{AB}$ that appears in formula \eqref{eq:DvNeq2BT}.

The Donaldson-Witten twist of $N=2$ 4-dimensional super Yang-Mills theory is also an example of an $N=1$ BT-type equivariant twist. The details are quite similar to those of the GL twist described above and are thus omitted.

\section*{Acknowledgements}
I would like to gratefully acknowledge initial collaboration with Leron Borsten, Dimitri Kanakaris, and Hyungrok Kim. Their work on the topic has now appeared on the arXiv \cite{Borsten:2025dyv}. I am also indebted to Andrei Mikhailov for correspondence and enlightening discussions.

 I am funded
by the Croatian Science Foundation (HRZZ) project ``QFIST  — Quantum Fields, Symmetries, and Strings''  (UIP-2025-02-8940).

\appendix

\section{On conventions for the complex conjugation of fermions}
\label{appendix:CC}
There are two such conventions. One is the typical physics convention, where conjugation reverses the order:
\be
\overline{\alpha \beta}=\bar\beta \bar\alpha = (-1)^{\alpha \beta} \bar\alpha \bar\beta\,.
\ee
(In the above formula, the expression $\alpha \beta$ in the exponent denotes the product of the total Grassmann parities of $\alpha$ and $\beta$ modulo 2.) The other is the order-preserving convention:
\be
(\alpha \beta)^\ast=\alpha^\ast \beta^\ast\,.
\ee

I want a recipe that transforms expressions such that
\begin{enumerate}
    \item any expression which is real in one convention is mapped to a real expression in the other, and also
    \item such that any expression annihilated by a supersymmetry generator in one convention is also annihilated by a (transformed) supersymmetry generator in the other convention.
\end{enumerate} 
To achieve 1., simply introduce the map
\be
I(\alpha)= i^{\frac{1}{2}(F(\alpha) (F(\alpha)-1))}\alpha
\ee
where $F(\alpha)$ counts the number of fundamental fermions in $\alpha$ and is valued in $\mathbb Z$ rather than $\mathbb Z \mod 2$. $I$ is a linear self-map of an exterior algebra over $\mathbb C$ and $F$ counts the rank of the form denoted by $\alpha$, and is extended to inhomogeneous forms by linearity. It is easy to check that
\be
I\alpha^\ast=\overline{I\alpha}
\ee
so that $\alpha=\alpha^\ast \iff I\alpha=\overline{I\alpha}$.

Unfortunately, this map is not a morphism: $I(\alpha \beta)=i^{F(\alpha)F(\beta)}\alpha \beta$. Therefore if $Q$ defines an infinitesimal supersymmetry, it is not the case that $I Q I^{-1}$ does, because it will not satisfy a graded Leibniz rule.

There is a remedy, however. If I call $F(\alpha)\equiv n$ for brevity, then $I(\alpha)=i^{\frac{1}{2}n(n-1)}\alpha$. The prefactor
\be
i^{\frac{1}{2}n(n-1)}
\ee
depends on $n\mod 8$, because the exponent is taken modulo $4$ ($i^4=1$). Therefore insofar as writing the action of $I$ is concerned one only really need consider $F(\alpha) \mod 8$.

Therefore, if $Q$ is a supersymmetry and $L$ is a real expression in one complex conjugation convention that has $QL=0$, we may split in fermion number $\mod 8$ as
\be
\begin{split}
    L=L_0 + L_2 + L_4 + L_6\\
    Q=Q_1+Q_3+Q_5+Q_7
\end{split}
\ee
and then $QL=0$ gives rise to four equations:
\be
\begin{split}
    Q_1 L_0 + Q_3 L_6 + Q_5 L_4 + Q_7 L_2=0\,,\\
    Q_1 L_2+ Q_3 L_0 + Q_5 L_6+ Q_7 L_4=0\,, \\
    \cdots
\end{split}
\ee

Denote $L$ and $Q$ in the new convention by $L'$ and $Q'$. By the above discussion we can set $L'= I^{-1} L$. It is now is easy to deduce how to rescale each $Q_m$ so that $Q'L'=0 \iff QL=0$. If we fix $Q_1'=Q_1$, then the correct replacements are
\be
Q_3'=-i Q_3\,,\quad Q_5'=-Q_5\,,\quad Q_7'= i Q_7
\ee
or 
\be
Q_m'= i^{-(m-1)/2} Q_m\,.
\ee

\let\oldbibliography\thebibliography
\renewcommand{\thebibliography}[1]{%
    \oldbibliography{#1}%
    \setlength{\itemsep}{-1pt}%
}
\begin{multicols}{2}
{\setstretch{0}
    \small
    \bibliography{Alex_notes.bib}}
\end{multicols}
\end{document}